\documentclass[12pt]{article}
\usepackage{epsfig,amssymb,amsmath,psfrag,subfigure,rotate,color,wasysym}

\allowdisplaybreaks
\newcommand{\insertfig}[2]{\includegraphics[width=#1cm]{#2}}

\def\XXint#1#2#3{{\setbox0=\hbox{$#1{#2#3}{\int}$ }
\vcenter{\hbox{$#2#3$ }}\kern-.6\wd0}}

\def \be  {\begin{equation}}
\def \ee  {\end{equation}}
\def \ba  {\begin{eqnarray}}
\def \ea  {\end{eqnarray}}
\def \baa {\begin{eqnarray*}}
\def \eaa {\end{eqnarray*}}
\def \lab #1 {\label{#1}}

\newcommand\re[1]{(\ref{#1})}
\def\d{\hbox{{d}\kern-.20em\hbox{l}}}

\def \matrix #1 {\left(\begin{array}{cc} #1 \end{array}\right)}

\newcommand \vev [1] {\langle{#1}\rangle}

\newcommand \ket [1] {|{#1}\rangle}

\newcommand{\bit}[1]{\mbox{\boldmath$#1$}}

\newcommand{\ft}[2]{{\textstyle\frac{#1}{#2}}}
\numberwithin{equation}{section}
\begin{document}

\begin{titlepage}

\thispagestyle{empty}

\vspace*{3cm}

\centerline{\large \bf Supersymmetric quantum mechanics of the flux tube}
\vspace*{1cm}

\centerline{\sc A.V.~Belitsky}

\vspace{10mm}

\centerline{\it Department of Physics, Arizona State University}
\centerline{\it Tempe, AZ 85287-1504, USA}

\vspace{2cm}

\centerline{\bf Abstract}

\vspace{5mm}

The Operator Product Expansion approach to scattering amplitudes in maximally supersymmetric gauge theory operates in terms of pentagon transitions 
for excitations propagating on a color flux tube. These obey a set of axioms which allow one to determine them to all orders in 't Hooft coupling and confront
against explicit calculations. One of the simplifying features of the formalism is the factorizability of multiparticle transitions in terms of single-particle ones. 
In this paper we extend an earlier consideration of a sector populated by one kind of excitations to the case of a system with fermionic as well as bosonic 
degrees of freedom to address the origin of the factorization. While the purely bosonic case was analyzed within an integrable noncompact open-spin chain 
model, the current case is solved in the framework of a supersymmetric sl$(2|1)$ magnet. We find the eigenfunctions for the multiparticle system making use 
of the R-matrix approach. Constructing resulting pentagon transitions, we prove their factorized form. The discussion corresponds to leading order of perturbation 
theory.

\end{titlepage}

\setcounter{footnote} 0

\newpage

\pagestyle{plain}
\setcounter{page} 1

{
\footnotesize 
\tableofcontents}

\newpage

\section{Introduction}

\begin{figure}[t]
\begin{center}
\mbox{
\begin{picture}(0,280)(80,0)
\put(0,-170){\insertfig{20}{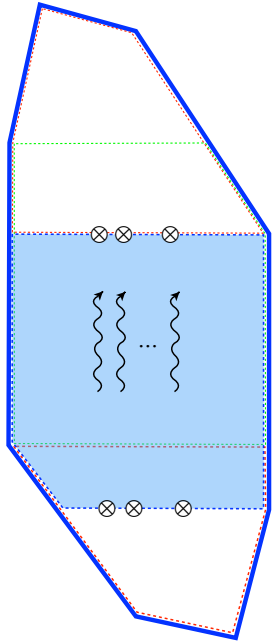}}
\end{picture}
}
\end{center}
\caption{\label{pentagonFig} Tessellation of a polygon into null squares merged into pentagons (shown in different dashed contours). We picked an intermediate
pentagon transition with flux-tube excitations inserted in the bottom and top portions of the contour.}
\end{figure}

The Operator Product Expansion for scattering amplitudes \cite{Basso:2013vsa,Alday:2010ku} of planar maximally supersymmetric Yang-Mills theory in the dual language of the 
Wilson loop stretched on a null polygonal contour in superspace \cite{Alday:2007hr,Drummond:2007cf,Brandhuber:2007yx,CaronHuot:2010ek,Mason:2010yk,Belitsky:2011zm} 
paved a way for their weak and strong coupling analysis in a multi-collinear limit with a naturally built-in consistent scheme for inclusion of subleading corrections 
\cite{Basso:2013aha,Belitsky:2014rba,Basso:2014koa,Belitsky:2014sla,Basso:2014nra,Belitsky:2014lta,Belitsky:2015efa,Basso:2014hfa,Basso:2014jfa,Basso:2015rta,%
Fioravanti:2015dma,Belitsky:2015qla,Bonini:2015lfr,Belitsky:2015lzw}. It is a based on geometrization of the contour in terms of a sequence of null squares with adjacent 
ones sharing a side merged into pentagons, see Fig.\ \ref{pentagonFig}. The bottom of the loop can be decomposed into an infinite series of excitations with the strength of  
contributions being exponentially suppressed with their number (or more precisely, their cumulative twist) in the collinear limit. These propagate upwards from the bottom through a 
series of pentagons and are absorbed at the top. Every pentagonal Wilson loop in the chain of transitions contains insertions of elementary fields of the theory with their total 
quantum numbers fixed by the choice of the component of the amplitude under study. These pentagons play a pivotal role in the entire construction. They obey a set of natural 
axioms \cite{Basso:2013vsa} that are inherited from the integrable dynamics of the $\mathcal{N}=4$ supersymmetric Yang-Mills theory. However, the question of their operatorial 
origin remains obscure. 

%%%%%%%%%%%%%%%%%%%%%%%%%%%%%%%%%%%%%%%%%%%%%%%%%%%%%%%%%%%%%%%%%%%%%
%            Figure
%%%%%%%%%%%%%%%%%%%%%%%%%%%%%%%%%%%%%%%%%%%%%%%%%%%%%%%%%%%%%%%%%%%%%
%%%%%%%%%%%%%%%%%%%%%%%%%%%%%%%%%%%%%%%%%%%%%%%%%%%%%%%%%%%%%%%%%%%%%

\subsection{Embedding of different multiplets}

Some time ago \cite{Belitsky:2014rba}, we studied the system of excitations of a single type interacting on the flux-tube. It was shown to be equivalent to solving the spectral
problem for a noncompact open spin chain whose sl(2) invariance is broken by boundary Hamiltonians. Presently we will provide its generalization to the minimally supersymmetric 
sector of the $\mathcal{N} = 4$ super Yang-Mills theory in the planar limit. In the absence of a covariant  superspace formulation of the theory, the light-cone formalism becomes 
advantageous. In this framework, all propagating fields in the maximally supersymmetric Yang-Mills theory can be accommodated into a single light-cone chiral superfield
\cite{Brink:1982pd,Mandelstam:1982cb,Belitsky:2004sc},
\begin{align}
\label{N4SYMsupefield}
\Phi_{\mathcal{N} = 4} (x^\mu, \theta^A) 
&= 
\partial_+^{-1} A (x^\mu) + \theta^A \partial_+^{- 1} \bar\psi_A (x^\mu) + \frac{i}{2!} \theta^A \theta^B \phi_{AB} (x^\mu)
\\
&
+
\frac{1}{3!} \varepsilon_{ABCD} \theta^A \theta^B \theta^C \psi^D (x^\mu) - \frac{1}{4!} \varepsilon_{ABCD} \theta^A \theta^B \theta^C \theta^D \partial_+ \bar{A} (x^\mu)
\, , \nonumber
\end{align}
where $A$ and $\bar{A}$ are the holomorphic and antiholomorphic components of the gauge field, respectively, $\psi$ and $\bar\psi$ are the dynamical
``good'' components of the fermion fields transforming in the ${\bf 4}$ and ${\bf \bar{4}}$ of the internal $SU(4)$ symmetry group and, finally, $\phi_{AB}$
is a sextet of scalars. 

There are two possible subsectors we can analyze. One them is of Wess-Zumino type. It is is composed of a scalar and a fermion
\begin{align}
\label{WZsuperfield}
\Phi_{s = 1/2} (x^\mu, \theta) = \phi (x^\mu) + \theta \psi (x^\mu)
\end{align}
and is obtained from \re{N4SYMsupefield} via the projection \cite{Belitsky:2007zp}
\begin{align}
\label{projectionWZmultiplet}
\Phi_{\mathcal{N} = 4} (x^\mu, \theta^A) |_{\theta^2 = \theta, \theta^3 = 0} = \dots + \theta^1 \theta^4 \Phi_{s = 1/2} (x^\mu, \theta)
\, . 
\end{align}
The other one is the antiholomorphic part of the $\mathcal{N} = 1$ superYang-Mills multiplet,
\begin{align}
\label{N1SYMmultiplet}
\Phi_{s = 1} (x^\mu, \theta) = \psi (x^\mu) - \theta \bar{F} (x^\mu)
\, ,
\end{align}
built from a fermion and antiholomorphic field strength $\bar{F} = \partial_+ \bar{A}$, found in the top two components of the $\mathcal{N} = 4$ superfield,
\begin{align}
\Phi_{\mathcal{N} = 4} (x^\mu, \theta^A) |_{\theta^4 = \theta} = \dots + \theta^1 \theta^2 \theta^3 \Phi_{s = 1} (x^\mu, \theta)
\end{align}
In both cases, we displayed the conformal spin of the minisuperfield, which is determined by the one of its lowest field component, as a subscript.

\subsection{Superlight-cone operators and Hamiltonians}
\label{MEhamiltonianSection}

As can be seen from the representation of the pentagon transition in Fig.\ \ref{pentagonFig}, it is related to the correlation function of two $\Pi$-shaped Wilson loops 
\cite{Belitsky:2011nn,Sever:2012qp} with insertions of elementary fields into their bottom and top contours, schematically
\begin{align}
P ({\rm bottom} | {\rm top}) \sim \vev{O_{\Pi_{\rm top}} \mathcal{O}_{\Pi_{\rm bottom}} }
\, ,
\end{align}
where 
\begin{align}
\label{PiLCOperator}
\mathcal{O}_{\Pi} (\bit{Z}) 
=
W^\dagger (0) \Phi_s (Z_1) \Phi_s (Z_2) \dots \Phi_s (Z_N) W (\infty)
\, ,
\end{align}
is built from superfields $\Phi$ inserted along the light-cone direction $z_n = x^-_n$ and depends on respective Grassmann variable $\theta_n$ that together can be 
encoded in a superspace coordinate $Z_n = (z_n, \theta_n)$. The gauge links between supercordinates $\bit{Z} = (Z_1, Z_2, \dots, Z_N)$ can be ignored due to the 
choice of the light-cone gauge condition $A^+ = 0$. The two light-like Wilson lines $W$ in the direction of particles propagating along the vertical segments of the pentagon 
are attached at its ends.

At leading order of perturbation theory (and multicolor limit), the renormalization group evolution of these operators can be cast in the form of a Sch{\"o}dinger 
equation with Hamiltonian given by the sum of pairwise Hamiltonians between adjacent superfields supplemented with the interaction of the first and last 
one with the boundary Wilson lines. The latter read
\begin{align}
\label{HamiltonianII}
\mathcal{H}_{01} W^\dagger (0) \Phi_s (Z_1) 
&
= W^\dagger (0)  \int_0^1 \frac{d \beta}{1 - \beta} \left[  \beta^{2 s - 1} \Phi_s (\beta Z_1) - \Phi_s (Z_1) \right]
\, , \\
\mathcal{H}_{N \infty} \Phi_s (Z_N ) W (\infty)
&
= \int_1^\infty \frac{d \beta}{\beta - 1} \left[ \Phi (\beta z_N, \theta_N) - \beta^{-1} \Phi_s (Z_N) \right] W (\infty) 
\, .
\end{align}

We can use the light-cone superspace formulation of the $\mathcal{N} = 4$ dilatation operator \cite{Belitsky:2004sc} to project out following Ref.\ \cite{Belitsky:2007zp} 
the scalar-fermion sector in question or directly get the $\mathcal{N}=1$ superYang-Mills \cite{Belitsky:2004sc} for the multiplet \re{N1SYMmultiplet}. The $\mathcal{N} = 4$ 
pair-wise Hamiltonian for superfields $\Phi_{\mathcal{N} = 4}$ of conformal spin $s = - \ft12$ sitting away from the boundary Wilson lines is 
\begin{align}
\mathcal{H}_{12} \Phi_{\mathcal{N} = 4} (Z_1) \Phi_{\mathcal{N} = 4} (Z_2)
&
=
\int_0^1 \frac{d \alpha}{\alpha}
\bigg[
(1 - \alpha)^{-2} \Phi_{\mathcal{N} = 4} ((1 - \alpha)Z_1 + \alpha Z_2) \Phi_{\mathcal{N} = 4} (Z_2)
\nonumber\\
&\qquad\qquad
+
(1 - \alpha)^{-2}  \Phi_{\mathcal{N} = 4} (Z_1) \Phi_{\mathcal{N} = 4} ((1 - \alpha)Z_2 + \alpha Z_1)
\nonumber\\
&\qquad\qquad
-
2 \Phi_{\mathcal{N} = 4} (Z_1) \Phi_{\mathcal{N} = 4} (Z_2)
\bigg]
\, . 
\end{align}
Projecting out the Wess-Zumino multiplet via \re{projectionWZmultiplet} changes the power of the $\alpha$-dependent prefactor from $-2$ to $0$. For the
antiholomorphic Yang-Mills multiplet \re{N1SYMmultiplet}, the same power changes from $-2$ to $1$. We can combine the two options by encoding 
them in the exponent $2s-1$. Let us change the integration variables in the integrand of $\mathcal{H}_{12}$, as well as modify the subtraction term, 
i.e., 
\begin{align}
\mathcal{H}^\prime_{12} = \mathcal{H}_{12} + \delta \mathcal{H}_{12} 
\, , \qquad
\mbox{with}
\qquad
\delta \mathcal{H}_{12} = \ln z_2/z_1
\, ,
\end{align}
such that in the limit $z_2 \gg z_1$, we get the sum of two boundary Hamiltonians \re{HamiltonianII}. Here, the pair-wise Hamiltonian is split in two
\begin{align}
\label{IntermediateH12}
\mathcal{H}^\prime_{12} \Phi_s (Z_1) \Phi_s (Z_2)
=
\mathcal{H}^+_{12} \Phi_s (Z_1) \Phi_s (Z_2)
+
\mathcal{H}^-_{12} \Phi_s (Z_1) \Phi_s (Z_2)
\, ,
\end{align}
that act in the following fashion on the nearest-neighbor fields
\begin{align}
\mathcal{H}^-_{12} \Phi_s (Z_1) \Phi_s (Z_2)
&
=\!
\int_1^{z_2/z_1}
\frac{d \beta}{\beta - 1}
\\
&
\times\!
\left[
\left(
\frac{z_2 - \beta z_1}{z_2 - z_1}
\right)^{2 s - 1}
\!\Phi_s\! \left( \beta z_1, \frac{z_2 - \beta z_1}{z_2 - z_1} \theta_1 + \frac{z_1 (\beta - 1)}{z_2 - z_1} \theta_2 \right)
-
\beta^{-1} \Phi_s (Z_1)
\right]
\Phi_s (Z_2)
\, , \nonumber\\
\mathcal{H}^+_{12} \Phi_s (Z_1) \Phi_s (Z_2)
&
=
\Phi_s (Z_1)
\int_{z_1/z_2}^1
\frac{d \beta}{1 - \beta }
\\
&
\times\!
\left[
\left(
\frac{z_1 - \beta z_2}{z_1 - z_2}
\right)^{2 s - 1}
\Phi_s \left( \beta z_2, \frac{\beta z_2 - z_1}{z_2 - z_1} \theta_2 + \frac{z_2 (1 - \beta)}{z_2 - z_1} \theta_1 \right)
-
\Phi_s (Z_2)
\right]
\, . \nonumber
\end{align}
Thus the Hamiltonian that we have to solve the eigensystem for is
\begin{align}
\label{TotalHamiltonian}
\mathcal{H}_N = \mathcal{H}_{01} + \mathcal{H}'_{12} + \dots +  \mathcal{H}'_{N-1,N}  + \mathcal{H}_{N\infty}
\, .
\end{align}
Depending on the conformal spin of the superfields, it encodes both the Wess-Zumino and Yang-Mills multiplets. Since the two differ only by the value of the spin, 
the following discussion will be done for arbitrary $s$. This also points out that, while the scalars and fermions carry the R-charge in the $\mathcal{N}=4$ theory, 
this spin model will not be able to accommodate for nontrivial rational prefactors that arise in the pentagon approach, otherwise, these would arise in the fermion-gluon 
sectors as well. However, the latter is free from these `complications' since the gluon is singlet with respect to SU(4). Thus, the leading order description within the 
supersymmettric lattice model will provide information on the dynamical portion of the pentagons only. The rational factors as well as helicity form factors stemming 
from crossing conditions will not be accounted for within the current formalism.

\section{Supersymmetric open spin chain}

The light-cone chiral superfield $\Phi_s (Z)$ defines an infinite-dimensional chiral representation $\mathcal{V}_s$ of the superconformal sl$(2|1)$ algebra labeled by 
the conformal spin $s$. The generators of the algebra are realized as first order differential operators in bosonic $z$ and fermionic $\theta$ variables 
\begin{align}
S^- 
&
= - \partial_z
\, ,  
&
S^+ 
&
= z^2 \partial_z + 2 z s + z \theta \partial_\theta
\, , 
&
S^0 
&
= z \partial_z + s + \ft12 \theta \partial_\theta
\, ,
&
B 
&
= \ft12 \theta \partial_\theta - s
\, , \nonumber\\
V^- 
&= \partial_\theta
\, , 
&
W^- 
&
= \theta \partial_z
\, ,
&
V^+ 
&
= z \partial_\theta
\, , 
&
W^+ 
&
= \theta (z \partial_z + 2 s)
\, .
\label{GeneratorsSL21diffrep}
\end{align}
Thus, the Hamiltonian \re{TotalHamiltonian} defines a non-periodic homogeneous open superspin chain. We will demonstrate below
that it is in fact integrable.

\subsection{Scalar product and involution properties of generators}

As will be clear from our discussion it will be indispensable to introduce an inner product on the space of functions depending on superspace variable
$Z$. While the bosonic variable lives on the real axis, it is instructive to address the spectral problem by promoting it to the upper half of the complex 
plane. This formulation is of paramount importance for the construction of eigenfunctions (holomorphic functions in the upper semiplane) in the representation 
of Separated Variables \cite{Sklyanin:1995bm,Derkachov:2002tf,Belitsky:2014rba} and computation of various inner product \cite{Belitsky:2014rba}. The 
flux-tube matrix elements entering the Operator Product Expansion can be regarded as their boundary values.

The chiral scalar product on the space of superfunctions
\begin{align}
\bit{\Phi}_s (Z) = \Phi_s (z) + \theta \Phi_{s + 1/2} (z)
\, ,
\end{align}
holomorphic in the upper semiplane of the complex plane is defined as
\begin{align}
\label{sl21innerproduct}
\vev{\bit{\Phi}'_s|\bit{\Phi}_s}
=
\int [D Z]_s \, \big( \bit{\Phi}'_s (Z) \big)^\ast \bit{\Phi}_s (Z)
\, ,
\end{align}
where the sl$(2|1)$ invariant measure reads
\begin{align}
\label{sl21measure}
\int [D Z]_s 
= 
\frac{{\rm e}^{- i \pi (s - 1)}}{\pi}
\int d \theta^\ast d \theta
\int_{\Im{\rm m}[z] > 0} d z^\ast dz \, \left( z - z^\ast + \theta \theta^\ast \right)^{2 s - 1} 
\, .
\end{align}
Notice that the phases chosen in this inner product are correlated with the integration and involution rules adopted for Grassmann variables. Throughout the paper they obey the 
following rules 
\begin{align}
\label{GrassmannInvolution}
\int d \theta \, \theta = 1
\, , \qquad
\big( \theta' \theta \big)^\ast = \theta'^\ast \theta^\ast
\, .
\end{align}
In the component form, we find
\begin{align}
\label{sl21InnerInComponents}
\vev{\bit{\Phi}'_s | \bit{\Phi}_s}
=
 \int [Dz]_{s} \big( \Phi'_s (z) \big)^\ast \Phi_s (z)
+
\frac{1}{2 i s}
\int [D z]_{s + 1/2}  \big( \Phi'_{s+1/2} (z) \big)^\ast \Phi_{s+1/2} (z)
\, ,
\end{align}
where, e.g., $\Phi_s = \phi$ and $\Phi_{s + 1/2} = \psi$ for the field content of the $s=1/2$ multiplet \re{WZsuperfield}. Here we recognize in the first term the well-known expression 
for the bosonic sl$(2)$-invariant inner product with the measure
\begin{align}
\int [Dz]_{s} \equiv
(2 s - 1)
\frac{{\rm e}^{- i \pi (s - 1)}}{\pi} 
\int_{\Im{\rm m}[z] > 0} d z^\ast dz \, (z - z^\ast)^{2s - 2} 
\, .
\end{align}
Notice an extra phase in front of the second term in Eq.\ \re{sl21InnerInComponents} to make it real by virtue of Eq.\ \re{GrassmannInvolution} for fermionic fields.
For $s = 1$, one has to change $\phi \to \psi$ and $\psi \to - \bar{F}$. Since the resulting superfield \re{N1SYMmultiplet} is fermionic, one has to multiply the inner
product $\vev{\Phi'_s | \Phi_s}$ by an $i$ such that this phase will migrate from the second term to the first. We will imply this convention from now on so that we 
could avoid repetitive formulas corresponding to each case. This nuisance will not affect any of our considerations which follow.

We conventionally define the adjoint operator with respect to the inner product \re{sl21innerproduct} as
\begin{align}
\vev{\bit{\Phi}' | G \bit{\Phi}} = \vev{G^\dagger \bit{\Phi}' | \bit{\Phi}}
\, .
\end{align}
Then we can easily verify the following conjugation properties of the sl$(2|1)$ generators \re{GeneratorsSL21diffrep} using integration by parts
\begin{align}
\label{ConjugationGenerators}
\left( S^{\pm,0} \right)^\dagger 
= -  S^{\pm,0}
\, , \qquad
B^\dagger 
= 
B
\, , \qquad
\left(
V^\pm
\right)^\dagger
&
=
-
W^\pm
\, .
\end{align}
Notice that the chirality generator is hermtitian compared to antihermitian generators of the sl(2) subalgebra. From the involution rules for Grassmann variables, it
follows that
\begin{align}
\left( G G' \right)^\dagger = (-1)^{{\rm grad} G \, {\rm grad} G'} G'^\dagger G^\dagger
\, .
\end{align}

The Hilbert space of the $N$-site model spanned on the light-cone operators \re{PiLCOperator} is formed by the tensor product of Hilbert spaces at the position of each 
superfield $\otimes_{k=1}^N \mathcal{V}_{s, k}$. Then, one can immediately proof the self-adjoint property of the Hamiltonian \re{TotalHamiltonian},
\begin{align}
\mathcal{H}^\dagger_N = \mathcal{H}_N
\end{align}
with respect to the inner product for multivariable $\bit{Z} = (Z_1, Z_2, \dots, Z_N)$ function $\bit{\Phi}_s = \bit{\Phi}_s (\bit{Z})$,
\begin{align*}
\vev{\bit{\Phi}'_s|\bit{\Phi}_s}
=
\int \prod_{n = 1}^N [D Z_n]_s \, \big( \bit{\Phi}'_s (\bit{Z}) \big)^\ast \bit{\Phi}_s (\bit{Z})
\, .
\end{align*}
To see this more efficiently, it is convenient to recast the individual pair-wise Hamiltonians in the non-local form, 
\begin{align}
\label{BoundaryDiffH}
\mathcal{H}_{01} 
&
= \psi (1) - \psi (z_1 \partial_{z_1} + \theta_1 \partial_{\theta_1} + 2s)
\, , \\
\mathcal{H}_{N\infty} 
&
= \psi (1) - \psi (- z_N \partial_{z_N})
\, , 
\end{align}
for the boundary and
\begin{align}
\label{BulkDiffH}
\mathcal{H}_{12} 
&
= 2 \psi (1) - \psi (z_{12} \partial_{z_1} + \theta_{12} \partial_{\theta_1}  + 2s) - \psi (z_{21} \partial_{z_2} + \theta_{21} \partial_{\theta_2}  + 2s)
\, , \\
\delta \mathcal{H}_{12}
&
=
\ln z_2/z_1
\, ,
\end{align}
for bulk ones, respectively. In fact, we can rearrange different contributions entering the bulk into the boundary Hamiltonians to better match 
them to the ones emerging from R-matrices. Namely, splitting the logarithmic terms in $\delta \mathcal{H}_{12}$, we can identify the bulk Hamiltonian 
with the sl(2$|$1) invariant one $h_{12} = \mathcal{H}_{12}$, 
\begin{align}
h_{12}^- =  \psi (1) - \psi (z_{12} \partial_{z_1} + \theta_{12} \partial_{\theta_1}  + 2s) 
\, , \qquad
h_{12}^+ =  \psi (1) - \psi (z_{21} \partial_{z_2} + \theta_{21} \partial_{\theta_2}  + 2s) 
\, ,
\end{align}
while the boundary ones now read
\begin{align}
h_{01} = \mathcal{H}_{10} - \ln z_1 = - \ln \left(z_1^2 \partial_{z_1} + z_1 \theta_1 \partial_{\theta_1} + 2 s z_1 \right)
\, , \qquad
h_{N \infty} = \mathcal{H}_{N\infty} + \ln z_N = - \ln \partial_{z_N} 
\, .
\end{align}

\subsection{Integrals of motion and hermiticity issues}

Let us construct the integrals of motion of the $N$-site Hamiltonian following the standard procedure of the R-matrix approach \cite{TakFad79}.
The sl$(2|1)$ Lax operator acting on the direct product of a graded three-dimensional space and the chiral space $\mathcal{V}_{s,k}$
of the $k$-th site reads
\begin{align}
\mathbb{L}_k (u) 
= 
\left(
\begin{array}{ccc}
u + S^0_k - B_k & - W^-_k & S^-_k \\
- V^+_k & u - 2 B_k & V^-_k \\
S^+_k & - W^+_k & u - S^0_k - B_k
\end{array}
\right)
\, .
\end{align}
It depends on the complex spectral parameter $u$. Notice that for a generic case, the representation of the algebra is parametrized by the conformal
spin $s$ and chirality $|b| \neq s$. Thus the Lax operator can be viewed as a function of three linear combinations of $u$, $s$ and $b$, namely,
\begin{align}
\label{GenericLaxParameters}
u_1  = u + s - b
\, , \qquad
u_2 = u - 2 b
\, , \qquad
u_3  = u - s - b
\, ,
\end{align}
such that $\mathbb{L} (u) = \mathbb{L} (u_1, u_2, u_3)$ is a function of $u_\alpha$. For the chiral case at hand, $b = - s$. However, we will use three distinct 
$u_\alpha$ parameters below to our advantage. The product of $N$ of these (with the increasing site number from left to right) defines the monodromy matrix
\begin{align}
\label{MonodromyTN}
\mathbb{T}_N (u) 
=
\mathbb{L}_1 (u) \mathbb{L}_2 (u) \dots \mathbb{L}_N (u) 
=
\left(
\begin{array}{cc}
A_N^{[2]\times[2]} (u) & B_N^{[2]} (u) \\
C_N^{[2]} (u) & D_N (u) 
\end{array}
\right)
\, ,
\end{align}
where we displayed the dimensions of the corresponding blocks as superscripts, e.g., $B_N^{[2]} = (B_N^1, B_N^2)$ etc. Our focus will be on the element $D_N (u)$. 
As can easily be found from the Yang-Baxter equation, $D_N (u)$ commutes with itself for arbitrary values of the spectral parameter $[D_N (u'), D_N (u)] = 0$. And as
it will be established in the next section, it commutes with the Hamiltonian as well,
\begin{align}
\label{DHcommutativity}
[D_N (u), \mathcal{H}_N] = 0
\, .
\end{align}
$D_N (u)$ thus generates a family of commuting charges which arise as coefficients of degree $N$ polynomial in $u$. However, we immediately find ourselves
in a predicament, since the operator $i^N D_N (i w)$ is not self-adjoint! It is obvious already for one site, where the only charge ${d}_1$ reads
\begin{align}
\label{D1}
D_1 (u) = u + {d}_1 \, , \qquad  {d}_1 =  - S^0_1 - B_1
\, ,
\end{align}
with $S^0_1$ and $B_1$ having opposite conjugation properties in light of Eqs.\ \re{ConjugationGenerators}. This implies that the eigenvalues of the 
operator $D_N$ are not real. This is not a problem by itself, however, it implies that the Hamiltonian will share only a subset of the eigenfunctions of the 
latter, i.e., the ones that yield its real eigenvalues. In fact, the complex nature of $D_N$ eigenvalues will be a virtue rather than a bug explaining the incremental 
shift in energy eigenvalues for excitations propagating on the flux tube. One can always define a new self-adjoint operator
\begin{align}
\label{OmegaN}
\Omega_N (w) = i^N D_N ( i w) +  (- i)^N D^\dagger_N (- i w)
\, ,
\end{align}
that will possess real eigenvalues. However, since we will be devising a procedure to calculate the eigenstates of the Hamiltonian $\mathcal{H}_N$ based on a 
recursion for $D_N$, using $\Omega_N$ for this purpose will be a significant obstacle on this route.

\subsection{Commutativity}
\label{CommutativitySection}

For one- and two-site cases, the proof of commutativity can be done by brute force. Namely, for $N=1$, the Hamiltonian \re{TotalHamiltonian} in the representation 
\re{BoundaryDiffH} can be rewritten in terms of generators as
\begin{align}
\mathcal{H}_1 
=
2 \psi (1) - \psi \left( S^0_1 + B_1 + 2s \right) - \psi \left( - S^0_1 + B_1 + 2s \right)
\, .
\end{align}
It is obviously self-adjoint and commutes with \re{D1} by virtue of the sl$(2|1)$ commutator algebra. For $N=2$, the operator $D_2 (u)$ is a second order
polynomial in spectral parameter
\begin{align}
D_2 (u) = u^2 + u \, {d}_1+ {d}_2
\, ,
\end{align}
with operator coefficients
\begin{align}
{d}_1 
= - S^0_1 - S^0_2 - B_1 - B_2
\, , \qquad
{d}_2 
= 
S_1^+ S_2^- + (S_1^0 + B_1) (S_2^0 + B_2) - W_1^+ V_2^-
\, .
\end{align}
While the commutativity of $\mathcal{H}_2$ with ${d}_1$ is almost obvious, the same property for the second-order differential operator ${d}_2 $ is far from this. 
In fact, the direct calculation results in the following relations for individual components of the two-site Hamiltonian,
\begin{align}
[{d}_2, \mathcal{H}_{12}] 
&= - z_1 \partial_{z_1} - \theta_1 \partial_{\theta_1} + z_2 \partial_{z_2} + \theta_2 \partial_{\theta_2}
\, , \\
[{d}_2, \mathcal{H}_{01}] 
&= - z_1 \partial_{z_2} - \theta_1 \partial_{\theta_2} 
\, , \\
[{d}_2, \mathcal{H}_{2\infty}] 
&=
\frac{z_1}{z_2}
\left(
z_1 \partial_{z_1} + 2 s + \theta_1 \partial_{\theta_1} 
\right)
\, , \\
[{d}_2, \delta \mathcal{H}_{12}] 
&=
-
\frac{z_1}{z_2}
\left(
z_1 \partial_{z_1} + 2 s + \theta_1 \partial_{\theta_1} 
\right)
+
z_1 \partial_{z_2} + \theta_1 \partial_{\theta_2}
\\
&\ \ \ \ 
+
z_1 \partial_{z_1} + \theta_1 \partial_{\theta_1}
-
z_2 \partial_{z_2} + \theta_2 \partial_{\theta_2}
\, . \nonumber
\end{align}
Adding these together, one recovers the anticipated result \re{DHcommutativity} for $N=2$.

Beyond $N=2$, the direct proof becomes tedious and it is instructive to rely on the power of the R-matrix approach. In fact, as was demonstrated in
the seminal paper \cite{Derkachov:2005hw}, the $\mathcal{R}$-operator obeying the conventional Yang-Baxter relation
\begin{align}
\check{\mathcal{R}}_{12} (v - u) \mathbb{L}_1 (u) \mathbb{L}_2 (v) 
=
\mathbb{L}_1 (v) \mathbb{L}_2 (u) \check{\mathcal{R}}_{12} (v - u) 
\end{align}
with $\check{\mathcal{R}}_{12} = \Pi_{12} \mathcal{R}_{12}$ having the two quantum spaces intechanged with the permutation $\Pi_{12}$, can be 
factorized in terms of three intertwiners
\begin{align}
\check{\mathcal{R}}_{12} (v - u)
=
\mathcal{R}^{(1)}_{12} (v_1 - u_1)
\mathcal{R}^{(2)}_{12} (v_2 - u_2)
\mathcal{R}^{(3)}_{12} (v_3 - u_3)
\, ,
\end{align}
each exchanging only a pair of combinations of spectral parameters introduced in Eq.\ \re{GenericLaxParameters}, e.g.,
\begin{align}
\mathcal{R}^{(1)}_{12} (v_1 - u_1) \mathbb{L}_1 (v_1, u_2, u_3)  \mathbb{L}_2 (u_1, u_2, u_3) 
&=
\mathbb{L}_1 (v_1, u_2, u_3)  \mathbb{L}_2 (u_1, u_2, u_3) \mathcal{R}^{(1)}_{12} (v_1 - u_1)
\, , \\
\mathcal{R}^{(3)}_{12} (v_3 - u_3) \mathbb{L}_1 (v_1, v_2, v_3)  \mathbb{L}_2 (v_1, v_2, u_3) 
&=
\mathbb{L}_1 (v_1, v_2, u_3)  \mathbb{L}_2 (v_1, v_2, v_3) \mathcal{R}^{(3)}_{12} (v_3 - u_3)
\, ,
\end{align}
where\footnote{The consideration $\mathcal{R}^{(2)}$ was also done in \cite{Belitsky:2006cp}, however, it will not play any role in our construction and is thus completely 
disregarded.}
\begin{align}
\mathcal{R}^{(1)}_{12} (v_1 - u_1) \equiv \mathcal{R}^{(1)}_{12} (v_1| u_1, u_2, u_3)
\, , \qquad
\mathcal{R}^{(3)}_{12} (v_1 - u_1) \equiv \mathcal{R}^{(3)}_{12} (v_1, v_2, v_3| u_3)
\end{align}
depend in a translation invariant manner on the displayed spectral parameters and are actually independent of the ones not shown. They thus solve simplified $RLL$ 
relations displayed above.

For now, we will focus on $\mathcal{R}^{(3)}_{12}$ which is the generator of the bulk Hamiltonians. Namely, making use of its chiral limit $\mathbb{R}^{(3)}_{12}$ from the generic 
form derived in Ref.\ \cite{Belitsky:2006cp}, we find the following integral in the upper half of the complex plane that can be easily converted into the line integral 
representation for the function $\bit{\Phi}(Y_1, Y_2)$ of $Y_n = (y_n, \vartheta_n)$,
\begin{align}
\mathbb{R}^{(3)}_{12} (u) \bit{\Phi} (Y_1, Y_2)
&
=
\int [DZ]_s (y_1 - z^\ast + \vartheta_1 \theta^\ast)^{- u - 2 s} (y_2 - z^\ast + \vartheta_2 \theta^\ast)^{u} \bit{\Phi} (Z, Y_2)
\nonumber\\
&
=
\frac{\Gamma (2 s)}{\Gamma (- u) \Gamma (u+ 2s)}
\int_0^1 d \tau \, \tau^{- u - 1} \bar\tau^{u + 2 s - 1} 
\bit{\Phi} (\bar\tau Y_1 + \tau Y_2 , Y_2)
\, .
\end{align}
As can be easily verified expanding $\mathbb{R}^{(3)}_{12} (u)$ in the vicinity of $u = \varepsilon \to 0$, we find the Hamiltonian $h^-_{12}$,
\begin{align}
\mathbb{R}^{(3)}_{12} (\varepsilon) = 1 - \varepsilon h^-_{12} + O (\varepsilon^2)
\, ,
\end{align}
such that the $RLL$ relation to this order yields the commutation
\begin{align}
[h^-_{12}, \mathbb{L}_1 (v) \mathbb{L}_2 (v)] =  \mathbb{M}^-_1 \mathbb{L}_2 (v) -  \mathbb{L}_1 (v) \mathbb{M}^-_2 
\, , 
\end{align}
where
\begin{align}
\mathbb{M}^-_n 
= 
\left(
\begin{array}{ccc}
0 & 0 & 0 \\
0 & 0 & 0 \\
- z_n & \theta_n & 1
\end{array}
\right)
\, .
\end{align}
Similar relations can be found for $h^+_{12}$ by expanding in the vicinity of $u = - 2s + \varepsilon$ as $\varepsilon \to 0$. As one can see,
the bulk Hamiltonians commute with the monodromy matrix $\mathbb{T}_N$ \re{MonodromyTN} up to boundary terms. The latter are cancelled
by the boundary Hamiltonians $h_{01}$ and $h_{N\infty}$ in the same fashion as in the sl(2) case analyzed in \cite{Belitsky:2014rba}.

\section{Brute force diagonalization}

In this introductory section, we will perform the diagonalization of the Hamiltonian by solving the emerging differential equations for eigenfunctions of the generating 
function $D_N$ of conserved charges. We define an energy eigenstate of the flux-tube $\ket{E (\bit{\lambda})}$ with $N$ excitations possessing rapidities 
$\bit{\lambda} = (\lambda_1, \dots, \lambda_N)$. Then, the matrix element of the light-cone operator between the vacuum $\ket{0}$ and $\ket{E (\bit{\lambda})}$
\begin{align}
\label{GenericMatrixElement}
\bit{\Phi}_s (\bit{Z}; \bit{\lambda}) = \vev{ 0 | \mathcal{O}_{\Pi} (\bit{Z}) | E (\bit{\lambda}) }
\end{align}
will be an eigenfunction of $\mathcal{H}_N$.

\subsection{One-particle matrix element}

The solution of the one-particle problem is trivial as it arises from the first-order differential equation determining the eigensystem for $D_1$
\begin{align}
D_1 (i w) \bit{\Phi}_s (Z_1; \lambda_1) = (i w - i \lambda_1 + s) \Phi_s (z_1; \lambda_1) +  (i w - i \lambda_1 + s - \ft{1}{2}) \theta_1 \Phi_{s + 1/2} (z_1; \lambda_1)
\, .
\end{align}
It yields for the individual eigenfunctions
\begin{align}
\Phi_s (z_1; \lambda_1) = z_1^{i \lambda_1 - s}
\, , \qquad
\Phi_{s + 1/2} (z_1; \lambda_1) = z_1^{i \lambda_1 - s - 1/2}
\, ,
\end{align}
that define the one-superparticle matrix element
\begin{align}
\label{1PmatrixElement}
\bit{\Phi}_s (Z_1; \lambda_1) = \Phi_s (z_1; \lambda_1) + \theta_1 \Phi_{s + 1/2} (z_1; \lambda_1) 
\, .
\end{align}
These are plane wave with complex wave numbers. As we alluded to above, the eigenvalues of $D$ are complex. However, its eigenfunctions generate eigenvalues 
of the flux-tube Hamiltonian
\begin{align}
\mathcal{H}_1 \bit{\Phi}_s (Z_1; \lambda_1) = E_s (\lambda_1) \Phi_s (z_1; \lambda_1) + E_{s + 1/2} (\lambda_1) \theta_1 \Phi_{s + 1/2} (z_1; \lambda_1)
\, ,
\end{align}
with the well-known (one-loop) energy
\begin{align}
E_s (\lambda_1) =  2 \psi (1) - \psi (s - i \lambda_1) - \psi (s + i \lambda_1)
\, .
\end{align}

\subsection{Two-particle matrix element}

Now, we move on to the two-particle case. We decompose the eigenfunction of $D_2$ in double Grassmann series over the two fermionic variables
\begin{align}
\bit{\Phi}_s (\bit{Z}) = \Phi_{ss} (\bit{z}) + \theta_1 \Phi_{s+1/2, s} (\bit{z}) + \theta_2 \Phi_{s, s+1/2} (\bit{z}) + \theta_1 \theta_2 \Phi_{s+1/2, s+1/2} (\bit{z}) 
\, ,
\end{align} 
with individual components depending on the bosonic variables $\bit{z} = (z_1, z_2)$. We have to solve the following equation in the component form
\begin{align}
D_2 (i w) \bit{\Phi}_s (\bit{Z}) 
&
=
(i w - i \lambda_1 + s) (i w - i \lambda_2 + s) \Phi_{ss} (\bit{z}) 
\nonumber\\
&
+  (i w - i \lambda_1 + s - \ft12) (i w - i \lambda_2 + s) \left[  \theta_1 \Phi_{s+1/2,s} (\bit{z})  + \theta_2 \Phi_{s,s+1/2} (\bit{z}) \right] 
\nonumber\\
&
+ (i w - i \lambda_1 + s - \ft12) (i w - i \lambda_2 + s - \ft12) \theta_1 \theta_2 \Phi_{s+1/2,s+1/2} (\bit{z}) 
\, .
\end{align}
The first-order differential equations arising from it fix the overall plane-wave factors of various contributions. The second order differential equations determine the remaining 
function of the ratio $z_1/z_2$ accompanying the waves and read
\begin{align}
&
\left[ - z_1 (z_1 - z_2) \partial_{z_1} \partial_{z_2} - 2s \, z_1 \partial_{z_2} - (i \lambda_1 - s) (i \lambda_2 - s) \right] \Phi_{ss} (\bit{z}) = 0
\, , \\
&
\left[ - z_1 (z_1 - z_2) \partial_{z_1} \partial_{z_2} - 2 s \, z_1 \partial_{z_2} + z_1 \partial_{z_1} - (i \lambda_1 - s + \ft12) (i \lambda_2 - s) \right] \Phi_{s,s+1/2} (\bit{z}) = 0
\, , \\
\label{PsiOne}
&
\left[ - z_1 (z_1 - z_2) \partial_{z_1} \partial_{z_2} - ( (2s+1) z_1 - z_2) \partial_{z_2} - (i \lambda_1 - s + \ft12) (i \lambda_2 - s) \right] \Phi_{s+1/2,s} (\bit{z}) 
\nonumber\\
&\qquad\qquad\qquad\qquad\qquad\qquad\qquad\qquad
- (z_1 \partial_{z_1} + 2s) \Phi_{s, s+1/2} (\bit{z}) = 0
\, , \\
&
[ - z_1 (z_1 - z_2) \partial_{z_1} \partial_{z_2} - ( (2s+1) z_1 - z_2 ) \partial_{z_2} + z_1 \partial_{z_1} + 1 
\\
&\qquad\qquad\qquad\qquad\qquad\qquad\qquad\qquad
- 
(i \lambda_1 - s + \ft12) (i \lambda_2 - s + \ft12) ] \Phi_{s+1/2,s+1/2} (\bit{z}) = 0
\, . \nonumber
\end{align}
The solutions to these equations can be found in a straightforwards fashion
\begin{align}
\label{PhiSS}
\Phi_{ss} (\bit{z}) 
&= 
z_1^{i \lambda_1 - s} z_2^{i \lambda_2 - s} {_2F_1} \left. \left( {s + i \lambda_1, s - i \lambda_2 \atop 2 s} \right| 1 - \frac{z_1}{z_2} \right)
\, , \\
\label{Eigenfunctionpsi1}
\Phi_{s+1/2,s} (\bit{z})
&= 
(s + i \lambda_2)
z_1^{i \lambda_1 - s - 1/2} z_2^{i \lambda_2 - s} {_2F_1}
\left.\left(
{s + \ft12 + i \lambda_1, s - i \lambda_2 \atop 2 s + 1}
\right|
1 - \frac{z_1}{z_2}
\right)
\, , \\
\label{Eigenfunctionpsi2}
\Phi_{s, s+1/2} (\bit{z}) 
&= 
(s - i \lambda_2)
z_1^{i \lambda_2 - s} z_2^{i \lambda_1 - s - 1/2} {_2F_1}
\left.\left(
{s + i \lambda_2, s + \ft12 - i \lambda_1 \atop 2 s + 1}
\right|
1 - \frac{z_1}{z_2}
\right)
\, , \\
\label{PhiS/2S/2}
\Phi_{s + 1/2, s + 1/2} (\bit{z}) 
&= 
z_1^{i \lambda_1 - s - 1/2} z_2^{i \lambda_2 - s - 1/2} {_2F_1} \left. \left( {s + \ft12 + i \lambda_1, s + \ft12 - i \lambda_2 \atop 2 s + 1} \right| 1 - \frac{z_1}{z_2} \right)
\, .
\end{align}
Notice that the solution to  \re{PsiOne} is not unique since one can always add to it a solution of the homogeneous equation with an arbitrary coefficient! 
Particularly noteworthy is the following ones that solves Eq. \re{PsiOne}
\begin{align}
\Phi'_{s+1/2,s} (\bit{z}) 
&= 
- (s - i \lambda_2)
z_1^{i \lambda_1 - s - 1/2} z_2^{i \lambda_2 - s} {_2F_1} \left. \left( {s + 1 - i \lambda_2, s + 1/2 + i \lambda_1 \atop 2s + 1} \right| 1 - \frac{z_1}{z_2} \right)
\, ,
\end{align}
since it is given by a single hypergeometric function and thus can be cast in a concise ``pyramid'' representation to be introduced later. The difference between 
the two solutions $\Phi'_{s+1/2,s} - \Phi_{s+1/2,s}  $ is indeed a solution to the homogeneous equation. Finally for the mixed wave functions, there is yet another 
(trivial) solution to the eigenvalue equation for the $D_2$-operator, i.e., $\Phi_{s,s+1/2} = 0$, $\Phi_{s+1/2,s} \neq 0$, however, like the previous one, it does not 
lead to consistent eigenvalue equation for the Hamiltonian.

With above results in our hands, we can immediately verify that they yield correct eigenvalues of the Hamiltonian $\mathcal{H}_2$, namely, we find
\begin{align}
\label{2PHeigenvalue}
\mathcal{H}_2 \bit{\Phi}_s (\bit{Z} ; \bit{\lambda})
&
=
[E_s (\lambda_1) + E_s (\lambda_2)] \Phi_{ss} (\bit{z}) 
\nonumber\\
&
+
[E_{s+1/2} (\lambda_1) + E_{s+1/2} (\lambda_2)] \theta_1 \theta_2 \Phi_{s+1/2,s+1/2} (\bit{z}) 
\nonumber\\
&
+
[E_{s+1/2} (\lambda_1) + E_s (\lambda_2)] 
\left[
\theta_1 \Phi_{s+1/2,s} (\bit{z}) 
+
\theta_2 \Phi_{s,s+1/2} (\bit{z}) 
\right]
\, .
\end{align}
Notice that the two eigenfunctions $ \Phi_{s+1/2,s} (\bit{z}) $ and $ \Phi_{s,s+1/2} (\bit{z}) $ possess the same eigenvalue!

\section{Algebraic construction of eigenfunctions}
\label{AlgebraicEigenfunctionsSection}

Beyond $N=2$, i.e., for three sites and more, the brute force solution of higher-order differential equations is hopeless. Therefore, we will devise a
recursive algebraic procedure to find the eigenfunctions of the operator $D_N$. It will turn out that the formalism will produce only one representative 
solution at a given Grassmann degree. The rest however will be generated by means of supersymmetry. The procedure will be based on the intertwiner 
$\mathcal{R}^{(1)}_{12}$ introduced earlier in Section \ref{CommutativitySection} that will yield a closed recursion for the matrix element $D$ of the 
monodromy operator. However, we have to find first its representation on the space of chiral matrix elements.

\subsection{Lowest component}
\label{LowestComponentSection}

To start with let us recall the solution for the lowest component $\Phi_{s \dots s} (\bit{z})$ of the $N$-particle supermatrix element \re{GenericMatrixElement}.
It is determined by the sl(2) open spin chain that was addressed in Ref.\ \cite{Belitsky:2014rba}. The intertertwiner that is used in the recursive procedure to
solve the eigenvalue equation for the bosonic counterpart of $D_N$ reads \cite{Belitsky:2014rba}
\begin{align}
{\rm r}_{12}^{(1)} (u) {\Phi} (y_1, y_2)
&
=
\frac{\Gamma (2s + 1) \Gamma (y_{21} \partial_{y_2} + u + 2s)}{\Gamma (u + 2s + 1) \Gamma (y_{21} \partial_{y_2} + 2s)} {\Phi} (y_1, y_2)
\nonumber\\
&
=
\int [D z]_s (y_1 - z^*)^{u} (y_2 - z^*)^{- u - 2s} {\Phi} (y_1, z)
\, .
\end{align}
For instance, the two-particle eigenstate is
\begin{align}
\label{2Particle11}
\Phi_{ss} (\bit{z}; \bit{\lambda}) = z_1^{i \lambda_1 - s} {\rm r}_{12}^{(1)} ( - i \lambda_1 - s ) z_2^{i \lambda_2 - s}
=
z_1^{i \lambda_1 - s} z_2^{i \lambda_2 - s}
{_2 F_1} \left. \left( {s + i \lambda_1, s - i \lambda_2 \atop 2 s} \right| 1 - \frac{z_1}{z_2} \right)
\, ,
\end{align}
and agrees with Eq.\ \re{PhiSS} found earlier. For a generic $N$-particle case $\bit{z} = (z_1, z_2, \dots, z_N)$, we found
\begin{align}
\Phi_{s\dots s} (\bit{z}; \bit{\lambda}) 
=
z_1^{i \lambda_1 - s}
{\rm r}^{(1)}_{1 \dots N} (- i \lambda_1 - s) z_2^{i \lambda_2 - s} {\rm r}^{(1)}_{2 \dots N} (- i \lambda_2 - s) z_3^{i \lambda_3 - s}
\dots 
{\rm r}^{(1)}_{N-1N} (- i \lambda_N - s) z_N^{i \lambda_N - s}
\, ,
\end{align}
where
\begin{align}
{\rm r}^{(1)}_{n \dots N} (u)
=
{\rm r}^{(1)}_{N-1,N} (u)
{\rm r}^{(1)}_{N-2,N-1} (u)
\dots
{\rm r}^{(1)}_{n, n+1} (u)
\, .
\end{align}

Let us now turn to further components in the Grassmann expansion. To this end we have to deduce the intertwiner that will simplify the solution for
$D_N$ in the supersymmeric case.

\subsection{Chiral limit of factorized matrices}

The discussion in Ref.\ \cite{Belitsky:2006cp} was done for generic representations, i.e., involving both chiral and antichiral Grassmann variables $\theta$ and 
$\bar\theta$. We start therefore with derived there integral (zig-zag) representation for the intertwiner $\mathcal{R}^{(1)}$ and take its chiral limit.
Ignoring a convention-dependent normalization factor, we define
\begin{align}
\mathcal{R}_{12}^{(1)} (u) \bit{\Phi} (\mathcal{Y}_1, \mathcal{Y}_2)
=
\int [\mathcal{D} \mathcal{Z}]_{j \bar{j}} 
\mathcal{K}_{0, -u} (\mathcal{Y}_1, \mathcal{Z}^*) \mathcal{K}_{j, \bar{j} + u} (\mathcal{Y}_2, \mathcal{Z}^*)
\bit{\Phi} (\mathcal{Y}_1, \mathcal{Z})
\, ,
\end{align}
where $\mathcal{Y} = (y, \vartheta, \bar\vartheta)$, $\mathcal{Z} = (z, \theta, \bar\theta)$ and the measure reads
\begin{align}
\int [\mathcal{D} \mathcal{Z}]_{j \bar{j}}
=
\frac{j + \bar{j}}{j \bar{j}} \int_{\Im{\rm m}[z] > 0}  \frac{d^2 z}{\pi} \int d \theta d \theta^* \int d \bar\theta d \bar\theta^* \, 
(z_+ - z_+^* - \theta \theta^* )^{j} (z_- - z_-^* - \bar\theta \bar\theta^* )^{\bar{j}}
\end{align}
along with the reproducing kernel 
\begin{align}
\mathcal{K}_{j\bar{j}} (\mathcal{Y}, \mathcal{Z}^*) = (y_+ - z_+^* - \vartheta \theta^* )^{- j} (y_- - z_-^* - \bar\vartheta \bar\theta^* )^{- \bar{j}}
\, .
\end{align}
Here, we introduced a notation for (anti)chiral bosonic coordinates $z_\pm = z \pm \ft12 \bar\theta \theta$,  $y_\pm = y \pm \ft12 \bar\vartheta \vartheta$
with conjugate ones found according to the rule \re{GrassmannInvolution}.

To reach the chiral limit in the above expressions, we take into account that $\Phi$ depends on $\bar\theta$ only through the chiral bosonic coordinates,
\begin{align}
\bit{\Phi} (\mathcal{Y}_1, \mathcal{Y}_2) = \bit{\Phi} (Y_1, Y_2)
\, ,
\end{align}
with $Y = (y_+, \vartheta)$. Then, shifting the bosonic integration variables $z_+ \to z$, we can perform the integration with respect to $\bar\theta$
and $\bar\theta^*$ and send $\bar{j} \to 0$ and $j \to 2 s$ afterwards. We obtain 
\begin{align}
\mathbb{R}^{(1)}_{12} (u) \bit{\Phi} (Y_1, Y_2)
=
\int [D Z]_s (y_1 - z^* - \theta \theta^*)^u (y_2 - z^* - \theta \theta^*)^{- u} (y_2 - z^* - \vartheta_2 \theta^*)^{- 2 s} \bit{\Phi} (Y_1, Z) 
\, ,
\end{align}
where the chiral integration measure was introduced earlier in Eq.\ \re{sl21measure}. One can actually rewrite this operator in terms of a nonlocal differential 
operator using the properties of bosonic reproducing kernels such that it reads explicitly
\begin{align}
\mathbb{R}^{(1)}_{12} (u) = \frac{\Gamma (2s + 1) \Gamma (y_{21} \partial_{y_2} + u + 2s + 1)}{\Gamma (u + 2s + 1) \Gamma (y_{21} \partial_{y_2} + 2s + 1)}
\, .
\end{align}
In turn, it can be cast as an integral on the real line adopting the well-known integral representation for the Euler Beta function, or the bosonic integral in the 
upper half of the complex plain,
\begin{align}
\mathbb{R}_{12}^{(1)} (u) \bit{\Phi} (Y_1, Y_2)
&
=
\frac{\Gamma (2s + 1)}{\Gamma (- u)\Gamma (u + 2s + 1)} \int_0^1 d\tau \, \tau^{- u - 1} \bar{\tau}^{u + 2s} \bit{\Phi} (Y_1, \tau y_1 + \bar{\tau} y_2, \vartheta_2)
\nonumber\\
&
=
\int [D z]_{s + 1/2} (y_1 - z^*)^{u} (y_2 - z^*)^{- u - 2s - 1} \bit{\Phi} (Y_1, z, \vartheta_2)
\, .
\end{align}
Here, the conformal spin $s$ may be understood to admit two different values depending on the component field the operator it acts on, for instance for
$s \to s - 1/2$, we fall back into the bosonic case discussed in the earlier section, $\mathbb{R}_{12}^{(1)} (u)|_{s \to s - 1/2} = {\rm r}^{(1)}_{12} (u)$.

\subsection{Two-particle case}

Let us start applying above results to the derivation of the two-site eigenfunctions. Making use of the known right factorization property of the Lax operator, 
\begin{align}
\label{RightFactorizationL}
\mathbb{L}_n (v_1, u_2, u_3)
=
\mathbb{L}_n (u_1, u_2, u_3)
\mathbb{M}_n (u_1| v_1)
\, , \quad\mbox{with}\quad
\mathbb{M}_n (u_1| v_1)
=
\left(
\begin{array}{ccc}
v_1/u_1 & 0 & 0 \\
0             & 1 & 0 \\
(v_1/u_1-1) z_n & 0 & 1
\end{array}
\right)
\, ,
\end{align}
which allows us to restore the same spectral parameter in $\mathbb{L}_n$, we can write a relation between two- and one-site monodromy matrices \re{MonodromyTN}
\begin{align}
\mathbb{R}^{(1)}_{12} (v_1 - u_1) \mathbb{L}_1 (v_1, u_2, u_3) \mathbb{T}_{1} (u) = \mathbb{T}_2 (u) \mathbb{M}_2 (u_1 | v_1) \mathbb{R}^{(1)}_{12} (v_1 - u_1)
\, .
\end{align}
Projecting out the $33$-entry of the monodromy matrix $D_2 (u) = [\mathbb{T}_2 (u)]_{33}$ in the right-hand side, we end up with the relation
\begin{align}
\label{2partRecursion}
D_2 (u) \mathbb{R}_{12}^{(1)} (v_1 - u_1)
=
\mathbb{R}_{12}^{(1)} (v_1 - u_1)
\big[
&
z_1 (z_1 \partial_{z_1}  + \theta_1 \partial_{\theta_1} + v_1 - u_3) B^1_1 (u)
\\
-
&
\theta_1 (z_1 \partial_{z_1} + u_3 - u_2) B^2_1 (u)
+
(u_3 - z_1 \partial_{z_1} - \theta_1 \partial_{\theta_1}) D_1 (u)
\big]
\, , \nonumber
\end{align}
where the elements of the one-particle monodromy (i.e., the Lax operator itself) matrix in the right-hand side of this equation act only on the variables of the 
second site, i.e., 
\begin{align}
B_1^1 (u) = - \partial_{z_2}
\, , \qquad
B_1^2 (u) = \partial_{\theta_2}
\, , \qquad
D_1 (u)
=
u_3 - z_2 \partial_{z_2} - \theta_2 \partial_{\theta_2}
\, .
\end{align}

In order to construct a recursion, the first two terms in the right-hand side of Eq.\ \re{2partRecursion} have to vanish when acting on a state of our 
choice. There are two\footnote{In fact since $B_1^2 (u) = \partial_{\theta_2}$ is a derivative that annihilates a Grassmann constant, we can encode 
the lowest component into this bare function by adding a $\theta$-independent term $ \Phi^{(0)}_{s,s} (\bit{z})$. This eigenfunction was discussed in 
Section \ref{LowestComponentSection} already.} such choices cumulatively denoted by $\bit{\Phi}^{(0)} (\bit{Z})$,
\begin{align}
\bit{\Phi}^{(0)} (\bit{Z})
=
\theta_1 \Phi^{(0)}_{s+1/2,s} (\bit{z})
+
\theta_1 \theta_2 \Phi^{(0)}_{s+1/2,s+1/2} (\bit{z})
\, ,
\end{align}
where $\bit{Z} = (Z_1, Z_2)$ and $\bit{z} = (z_1, z_2)$. Notice that the Grassmann structure $\theta_2$ will necessarily involve $B$-operators and will not 
be closed under recursion. However, as will be demonstrated below, it can be found by virtue of supersymmetry. 

Since different degree Grassmann components do not talk to each other, we can analyze them separately. Let us start with the $\theta_1$ component and 
cast it in the factorized form $\theta_1 \Phi^{(0)}_{s+1/2,s} (\bit{z}) = \theta_1 z_1^\alpha \Phi^{(0)}_{s} (z_2) $ and fix the value of $\alpha$ from the vanishing 
of the action of the first term in the brackets, $\alpha = u_3 - v_1 - 1$ and provides the eigenvalue of the first level of recursion $v_1 = i w - i \lambda_1  - s + 
\ft12$.

\begin{align}
D_2 (u) \mathbb{R}_{12}^{(1)} (v_1 - u_1) \theta_1 \Phi^{(0)}_{s+1/2,s} (\bit{z}) 
=
(i w - i \lambda_1 - s + \ft12) \theta_1 z_1^{i \lambda_1 - s - 1/2} \mathbb{R}_{12}^{(1)} (v_1 - u_1) D_1 (u) z_2^\beta
\, ,  
\end{align}
such that $\alpha = i \lambda_1 - s - 1/2$ and $u_3 = i w$, $u_1 = u_2 = i w + s + 1/2$. Here we took into account that $\mathbb{R}_{12}^{(1)}$ acts on 
$z_2$ coordinate only such that we can move $z_1$-dependent factor to its left. Next, substituting $\beta = i \lambda_2 - s$ in $\Phi^{(0)}_{s} (z_2) = 
z_2^\beta$, we immediately obtain
\begin{align}
&
D_2 (i w) \mathbb{R}_{12}^{(1)}  ( - i \lambda_1 - s - \ft12) \Phi^{(0)}_{s+1/2,s} (\bit{z}) 
\\
&\qquad\quad
= ( i w - i \lambda_1 + s - \ft12) ( i w - i \lambda_2 + s ) z_1^{i \lambda_1 - s - 1/2} \mathbb{R}_{12}^{(1)} ( - i \lambda_1 - s - \ft12) z_2^{i \lambda_2 - s} 
\, , \nonumber
\end{align}
with the resulting eigenfunction being
\begin{align}
\label{2Particle21}
\Phi_{s+1/2,s} (\bit{z}; \bit{\lambda})
&
=
z_1^{i \lambda_1 - s - 1/2} \mathbb{R}_{12}^{(1)} (- i \lambda_1 - s - \ft12) z_2^{i \lambda_2 - s}
\\
&
=
z_1^{i \lambda_1 - s - 1/2} z_2^{i \lambda_2 - s}
{_2 F_1} \left. \left( {s + 1/2 + i \lambda_1, s - i \lambda_2 \atop 2 s + 1} \right| 1 - \frac{z_1}{z_2} \right)
\, . \nonumber
\end{align}
This, up to an overall normalization coefficient, is the result $\Phi_{s+1/2,s} (\bit{z})$ of the previous section. The missing prefactor that plays a crucial 
role in proper diagonalization of the Hamiltonian will be fixed making of supersymmetry later in this section.

For the highest component $\theta_1 \theta_2 \Phi^{(0)}_{s+1/2,s+1/2} (\bit{z}) $ adopting an analogous factorizable Ansatz $\Phi^{(0)}_{s+1/2,s+1/2} (\bit{z}) 
= z_1^\alpha  \Phi^{(0)}_{s+1/2}(z_2)$, with one-particle wave function $\Phi^{(0)}_{s+1/2}(z_2) = z_2^\beta$, we deduce in the same fashion
\begin{align}
&
D_2 (i w) \mathbb{R}_{12}^{(1)} (- i \lambda_1 - s - 1/2) \theta_1 \theta_2 \Phi^{(0)}_{s+1/2,s+1/2} (\bit{z}) 
\\
&\qquad\quad
=
(i w - i \lambda_1 + s - \ft12) (i w - i \lambda_2 + s - \ft12) \theta_1\theta_2 \Phi_{s+1/2,s+1/2} (\bit{z}) 
\, , \nonumber
\end{align}
with the explicit eigenfunction being
\begin{align}
\label{2Particle22}
\Phi_{s+1/2,s+1/2} (\bit{z}; \bit{\lambda}) 
&
=
z_1^{i \lambda_1 - s - 1/2} \mathbb{R}_{12}^{(1)} (- i \lambda_1 - s - 1/2) z_2^{i \lambda_2 - s - 1/2}
\\
&
=
z_1^{i \lambda_1 - s - 1/2} z_2^{i \lambda_2 - s - 1/2} {_2 F_1} \left. \left( {s + 1/2 + i \lambda_1, s + 1/2 - i \lambda_2 \atop 2 s + 1} \right| 1 - \frac{z_1}{z_2} \right)
\, . \nonumber
\end{align}

To summarize, the two-particle eigenfunctions constructed via the advocated algebraic procedure are (we added here the lowest component as well)
\begin{align}
\label{2ParticleEigenFunction}
\bit{\Phi}_s (\bit{Z}; \bit{\lambda})
=
\Phi_{ss} (\bit{z}; \bit{\lambda}) + \theta_1 \Phi_{s+1/2,s} (\bit{z}; \bit{\lambda}) + \theta_1 \theta_2 \Phi_{s+1/2,s + 1/2} (\bit{z}; \bit{\lambda})
\, ,
\end{align}
where the individual components are given by Eqs.\ \re{2Particle11}, \re{2Particle21} and \re{2Particle22}, respectively. Though, this construction does not allow one to 
find all eigenfunctions in the Grassmann expansion, e.g., in front of $\theta_2$ for the case at hand, and endow them with correct coefficients that they enter the supereigenfunction, 
one can use a recipe to restore all of them as suggested below.  

%%%%%%%%%%%%%%%%%%%%%%%%%%%%%%%%%%%%%%%%%%%%%%%%%%%%%%%%%%%%%%%%%%%%%
%            Figure
%%%%%%%%%%%%%%%%%%%%%%%%%%%%%%%%%%%%%%%%%%%%%%%%%%%%%%%%%%%%%%%%%%%%%
\begin{figure}[t]
\begin{center}
\mbox{
\begin{picture}(0,170)(240,0)
\put(0,-650){\insertfig{40}{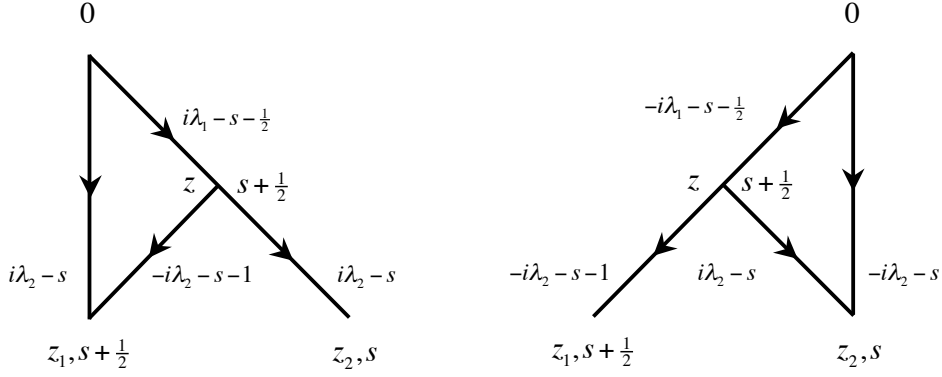}}
\end{picture}
}
\end{center}
\caption{\label{FigPyramid} Pyramid representation of the eigenfunction $\Phi_{s+1/2,s}$ (left panel) and its inverse (right panel).}
\end{figure}
%%%%%%%%%%%%%%%%%%%%%%%%%%%%%%%%%%%%%%%%%%%%%%%%%%%%%%%%%%%%%%%%%%%%%

Before, we outline it, let us introduce another representation for eigenfunctions which will be indispensable in the proof of their orthogonality
as well as analytic verification of factorizability of multiparticle pentagon transitions. It is the so-called pyramid representation which gives diagrammatic
interpretation for eigenfunctions in two-dimensional space. Making use of results in Appendix B of Ref.\ \cite{Belitsky:2014rba}, we can cast the above matrix elements in
the form
\begin{align}
\label{2PlowestPyramid}
\Phi_{ss} (\bit{z}; \bit{\lambda})
&=
z_1^{i \lambda_1 - s}
\int [D z]_{s} (z_1 - z^\ast)^{- i \lambda_1 - s} (z_2 - z^\ast)^{i \lambda_1 - s} z^{i \lambda_2 - s}
\, , \\
\label{2PhighestPyramid}
\Phi_{s+1/2,s+1/2} (\bit{z}; \bit{\lambda})
&=
z_1^{i \lambda_2 - s - 1/2}
\int [D z]_{s + 1/2} (z_1 - z^\ast)^{- i \lambda_2 - s - 1/2} (z_2 - z^\ast)^{i \lambda_2 - s - 1/2} z^{i \lambda_1 - s - 1/2}
\, ,
\end{align}
for the same-flavor components and
\begin{align}
\label{psi1One}
\Phi_{s+1/2,s} (\bit{z}; \bit{\lambda})
&=
z_1^{i \lambda_1 - s - 1/2}
\int [D z]_{s + 1/2} (z_1 - z^\ast)^{- i \lambda_1 - s - 1/2} (z_2 - z^\ast)^{i \lambda_1 - s - 1/2} z^{i \lambda_2 - s}
\\
&=
z_1^{i \lambda_2 - s}
\int [D z]_{s + 1/2} (z_1 - z^\ast)^{- i \lambda_2 - s - 1} (z_2 - z^\ast)^{i \lambda_2 - s} z^{i \lambda_1 - s - 1/2}
\, ,
\end{align}
for the $\theta_1$ component. Its graphical representation of (the second form of) this eigenfunction in terms of a ``pyramid'' is shown in Fig.\ \ref{FigPyramid}.
Now the missing eigenfunction can be simply found by promoting the internal bosonic propagators in the second representation 
to their supersymmetric extension
\begin{align}
\label{SuperPropagator}
(z' - z^\ast)^{- \alpha} \to [Z' - Z^\ast]^{- \alpha} \equiv (z' - z^\ast + \theta' \theta^\ast)^{- \alpha}
\, .
\end{align}
The Grassmann degree-one two-particle pyramid
\begin{align}
\Phi^{[1]}_2 (\bit{Z}; \bit{\lambda})
\equiv
\theta_1 \Phi_{s+1/2, s} (\bit{z}; \bit{\lambda}) 
+ 
\theta_2 \Phi_{s, s+1/2} (\bit{z}; \bit{\lambda})
\, ,
\end{align}
reads
\begin{align}
\label{2PmixedPyramid}
\Phi^{[1]}_2 (\bit{Z}; \bit{\lambda})
=
\int d \theta^\ast z_1^{i \lambda_2 - s} \int [D z]_{s + 1/2} [Z_1 - Z^\ast]^{- i \lambda_2 - s} [Z_2 - Z^\ast]^{i \lambda_2 - s} z^{i \lambda_1 - s - 1/2}
\, . \nonumber
\end{align}
Expanding the integrand in the fermionic variables, we uncover the missing solution $\Phi_{s, s+1/2}$ as well automatically produce the correct relative coefficients as 
functions of the rapidity variables.

\subsection{Three-particle case and beyond}

The one-third of the Yang-Baxter equation for the three-site case reads
\begin{align}
\mathbb{R}_{123}^{(1)} (v_1 - u_1) 
&
\mathbb{L}_1 (v_1, u_2, u_3)  \mathbb{L}_2 (u_1, u_2, u_3)  \mathbb{L}_3 (u_1, u_2, u_3) 
\nonumber\\
=
&
\mathbb{L}_1 (u_1, v_2, v_3)  \mathbb{L}_2 (u_1, u_2, u_3)  \mathbb{L}_3 (v_1, u_2, u_3) \mathbb{R}_{123}^{(1)} (v_1 - u_1)
\, ,
\end{align}
where 
\begin{align}
\mathbb{R}_{123}^{(1)} (v_1 - u_1) \equiv \mathbb{R}_{23}^{(1)} (v_1 - u_1)  \mathbb{R}_{12}^{(1)} (v_1 - u_1) 
\, .
\end{align}
Making use of Eq.\ \re{RightFactorizationL}, this relation can be rewritten for the momodromy matrices with decreasing number of sites
\begin{align}
\mathbb{R}_{123}^{(1)}  (v_1 - u_1)  \mathbb{L}_1 (v_1, u_2, u_3) \mathbb{T}_{2} (u) = \mathbb{T}_{3} (u) \mathbb{M}_3 (u_1 | v_1) \mathbb{R}_{123}^{(1)}  (v_1 - u_1) 
\, .
\end{align}
Extracting the $33$-matrix component from both sides and acting with the result on a test function $\Phi^{(0)} (\bit{Z})$ of three variables $\bit{Z} = (Z_1, Z_2, Z_3)$, 
we find
\begin{align}
\label{Diff3Particles}
\mathbb{R}_{123}^{(1)} (v_1 - u_1) 
\big[
&
z_1 (z_1 \partial_{z_1}  + \theta_1 \partial_{\theta_1} + v_1 - u_3) B^1_{2} (u)
-
\theta_1 (z_1 \partial_{z_1} + u_3 - u_2) B^2_{2} (u)
\nonumber\\
&
+
(u_3 - z_1 \partial_{z_1} - \theta_1 \partial_{\theta_1}) D_{2} (u)
\big]
\Phi^{(0)} (\bit{Z}) 
=
D_{3} (u) \mathbb{R}_{123}^{(1)} (v_1 - u_1)  \Phi^{(0)} (\bit{Z})
\, ,
\end{align}
where
\begin{align}
B_{2}^1 (u) 
&=  - (u_1 + z_2 \partial_{z_2}) \partial_{z_3} - \theta_2 \partial_{z_2} \partial_{\theta_3} - \partial_{z_2} (u_3 - z_3 \partial_{z_3} - \theta_3 \partial_{\theta_3})
\, , \\
B_{2}^2 (u)
&=  (z_2 \partial_{\theta_2} - (u_2 - u_1) \theta_2) \partial_{z_3} + (u_2 - \theta_2 \partial_{\theta_2} ) \partial_{z_3} + \partial_{\theta_2} (u_3 - z_3 \partial_{z_3} - \theta_3 \partial_{\theta_3})
\, .
\end{align}

To construct a self-contained recursion, we have to choose the bare three-particle wave function $\bit{\Phi}^{(0)} (\bit{Z})$ that eliminates the first two terms in Eq.\ \re{Diff3Particles}. 
It is achieved by the factorized Ansatz
\begin{align}
\bit{\Phi}^{(0)} (\bit{Z}) = \theta_1 z_1^{i \lambda_1 - s - 1/2} \bit{\Phi} (Z_2, Z_3)
\, ,
\end{align}
where we set $v_1 = i w - i \lambda_1$ and $\bit{\Phi} (Z_2, Z_3)$ is the two-particle eigenfunction whose three components were computed in the previous subsection. So the three of
the three-particle eigenfunctions are
\begin{align}
\bit{\Phi}_s (\bit{Z}) = \theta_1 z_1^{i \lambda_1 - s - 1/2} \mathbb{R}_{123}^{(1)} (- i \lambda_1 - s - \ft12) \bit{\Phi}_s (Z_2, Z_3)
\, ,
\end{align}
with $\bit{\Phi}_s (Z_2, Z_3)$ given in Eq.\ \re{2ParticleEigenFunction} with shifted labels of supercoordinates $k \to k+1$. Finally, the lowest, i.e., $\theta$-independent 
component of the eigenfunction can be found by eliminating any reference to Grassmann variables in the above equations, $\theta \to 0$, $\partial_\theta \to 0$, 
and was quoted in Section \ref{LowestComponentSection},
\begin{align}
\Phi_{sss} (\bit{z}) =  z_1^{i \lambda_1 - s} \mathbb{R}_{123}^{(1)} - i \lambda_1 - s) \Phi_{ss} (z_2, z_3)
\, ,
\end{align}
where the operator $\mathbb{R}_{123}^{(1)}$ is understood as the one with the shift in the spin $s \to s - 1/2$. Thus the vector of eigenfunction that can be obtained by means of the above 
algebraic constructions are
\begin{align}
\label{3ParticleEigenFunction}
\bit{\Phi}_s (\bit{Z}; \bit{\lambda}) 
&= 
z_1^{i \lambda_1 - s} \mathbb{R}_{123}^{(1)} z_2^{i \lambda_2 - s} \mathbb{R}_{23}^{(1)}  z_3^{i \lambda_3 - s} 
+ 
\theta_1
z_1^{i \lambda_1 - s - 1/2} \mathbb{R}_{123}^{(1)} z_2^{i \lambda_2 - s} \mathbb{R}_{23}^{(1)}  z_3^{i \lambda_3 - s} 
\\
&
+
\theta_1 \theta_2
z_1^{i \lambda_1 - s - 1/2} \mathbb{R}_{123}^{(1)} z_2^{i \lambda_2 - s - 1/2} \mathbb{R}_{23}^{(1)}  z_3^{i \lambda_3 - s} 
+
\theta_1 \theta_2 \theta_3
z_1^{i \lambda_1 - s - 1/2} \mathbb{R}_{123}^{(1)} z_2^{i \lambda_2 - s - 1/2} \mathbb{R}_{23}^{(1)}  z_3^{i \lambda_3 - s - 1/2} 
\, . \nonumber
\end{align}

The generalization to $N$-particle case is now straightforward,
\begin{align}
\label{NParticleEigenFunction}
\bit{\Phi}_s (\bit{Z}) 
&= 
z_1^{i \lambda_1 - s} \mathbb{R}_{1 2 \dots N}^{(1)} z_2^{i \lambda_2 - s} \mathbb{R}_{2 \dots N}^{(1)}  z_3^{i \lambda_3 - s} \dots \mathbb{R}_{N-1, N}^{(1)}  z_N^{i \lambda_N - s}
\\
&
+
\theta_1
z_1^{i \lambda_1 - s - 1/2} \mathbb{R}_{1 2 \dots N}^{(1)} z_2^{i \lambda_2 - s} \mathbb{R}_{2 \dots N}^{(1)}  z_3^{i \lambda_3 - s} \dots \mathbb{R}_{N-1, N}^{(1)}  z_N^{i \lambda_N - s}
\, , \dots \, , \ \nonumber\\
&
+
\theta_1 \theta_2 \dots \theta_N
z_1^{i \lambda_1 - s - 1/2} \mathbb{R}_{1 2 \dots N}^{(1)} z_2^{i \lambda_2 - s - 1/2} \mathbb{R}_{2 \dots N}^{(1)}  z_3^{i \lambda_3 - s - 1/2} \dots \mathbb{R}_{N-1, N}^{(1)}  z_N^{i \lambda_N - s - 1/2}
\, . \nonumber
\end{align}
Along the same route as was done in two-particle case, one can recover all eigenfunctions by employing supersymmetry at each level of odd variables in the pyramid representation 
of the eigenfunctions and, thus, restore relative coefficients accompanying them. 

\subsection{Orthogonality}
\label{OrthogonalitySection}

Before, we move on to using the above eigenfunctions for the calculation of pentagon transitions, let us prove their orthogonality first. In fact the technique that will be used for it here
is readily adoptable for the calculation of the latter as well.

\subsubsection{One site}

To keep track of different components in the Grassmann expansion, it is convenient to introduce a marker variable $\varepsilon$ via $\theta \to \varepsilon \theta$ for the
in-state eigenfunction and, correspondingly, $\varepsilon'$ for the out state. Then, using \re{sl21InnerInComponents}, we find
\begin{align}
\vev{\bit{\Phi}_s (\lambda'_1) | \bit{\Phi}_s (\lambda_1)}
=
\vev{\Phi_s (\lambda'_1) | \Phi_s (\lambda_1)} + \frac{\varepsilon' \varepsilon}{2 i s} \vev{\Phi_{s + 1/2} (\lambda'_1) | \Phi_{s + 1/2} (\lambda_1)}
\end{align}
where the component inner products
\begin{align}
\label{Component1PinnerProduct}
\vev{\Phi_s (\lambda'_1) | \Phi_s (\lambda_1)} 
=
2 \pi {\rm e}^{- \pi \lambda_1} \mu^{- 1}_s (\lambda_1) \delta (\lambda'_1 - \lambda_1)
\, ,
\end{align}
are expressed in terms of the measure
\begin{align}
\label{FTmeasureMUs}
\mu_s (\lambda)
=
\frac{\Gamma (s + i \lambda_1) \Gamma (s - i \lambda_1)}{\Gamma (2 s)}
\, ,
\end{align}
for the spin-$s$ flux-tube excitation. For $s = 1/2$, these reduce to the hole and fermion excitations for $(\varepsilon \varepsilon')^0$ and $(\varepsilon \varepsilon')^1$, respectively.
While for $s = 1$, they accommodate the fermion as the lowest  and the gauge field as the highest component of the $\mathcal{N} = 1$ gauge supermultiplet.

\subsubsection{Permutation identity in superspace}

%%%%%%%%%%%%%%%%%%%%%%%%%%%%%%%%%%%%%%%%%%%%%%%%%%%%%%%%%%%%%%%%%%%%%
%            Figure
%%%%%%%%%%%%%%%%%%%%%%%%%%%%%%%%%%%%%%%%%%%%%%%%%%%%%%%%%%%%%%%%%%%%%
\begin{figure}[t]
\begin{center}
\mbox{
\begin{picture}(0,150)(240,0)
\put(0,-590){\insertfig{35}{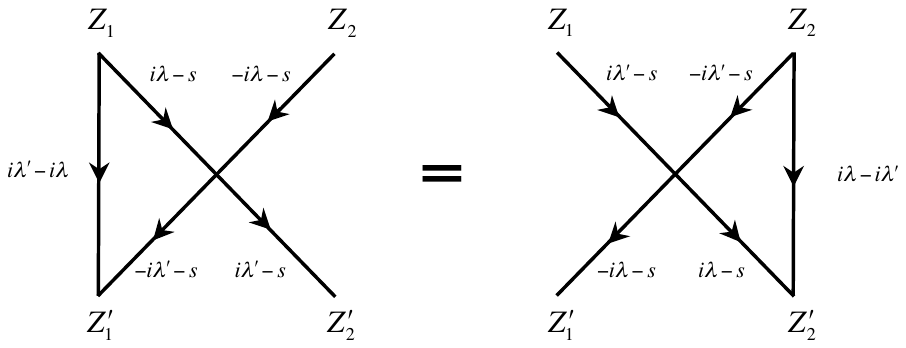}}
\end{picture}
}
\end{center}
\caption{\label{FigSuperpermutation} Superpermutation identity \re{superPermutationIdentity}.}
\end{figure}
%%%%%%%%%%%%%%%%%%%%%%%%%%%%%%%%%%%%%%%%%%%%%%%%%%%%%%%%%%%%%%%%%%%%%

To work out the two particle case and beyond, we have to introduce an identity that will be instrumental in the concise proof of orthogonality.
Namely, it is indispensable to use the permutation identity in the language of Feynman graphs lifted to the superspace. Introducing the superpropagator
\re{SuperPropagator}, from the superpoint $Z = (z, \theta)$ to $Z' = (z', \theta')$, one can show that
\begin{align}
\label{superPermutationIdentity}
[Z'_1 - Z_1^\ast]^{i \lambda' - i \lambda} \bit{X} \left( \bit{Z}; \lambda | \bit{Z}'; \lambda' \right)
=
\bit{X} \left( \bit{Z}; \lambda' | \bit{Z}'; \lambda \right) [Z'_2 - Z_2^\ast]^{i \lambda - i \lambda'}
\, ,
\end{align}
where the supercross is given by
\begin{align}
\bit{X} (\bit{Z}, \lambda | \bit{Z}', \lambda')
\equiv
\int [D Y]_s [Y - Z_1^\ast]^{i \lambda - s} [Y - Z_2^\ast]^{- i \lambda - s} [Z'_1 - Y^\ast]^{- i \lambda' - s} [Z'_2 - Y^\ast]^{i \lambda' - s}
\, .
\end{align}
It depends on four superpoints through $\bit{Z} = (Z_1, Z_2)$ and $\bit{Z}' = (Z'_1, Z'_2)$. Its form in terms of Feynman graphs is demonstrated in
Fig.\ \ref{FigSuperpermutation}. The identity reduces to its known bosonic counterpart, when all external Grassmann variables are set to zero, see 
Appendix \ref{FeynmanGraphsAppendix}.

\subsubsection{Two sites and more}
\label{SectionOrthogonalityPhi}

%%%%%%%%%%%%%%%%%%%%%%%%%%%%%%%%%%%%%%%%%%%%%%%%%%%%%%%%%%%%%%%%%%%%%
%            Figure
%%%%%%%%%%%%%%%%%%%%%%%%%%%%%%%%%%%%%%%%%%%%%%%%%%%%%%%%%%%%%%%%%%%%%
\begin{figure}[t]
\begin{center}
\mbox{
\begin{picture}(0,220)(270,0)
\put(0,-430){\insertfig{30}{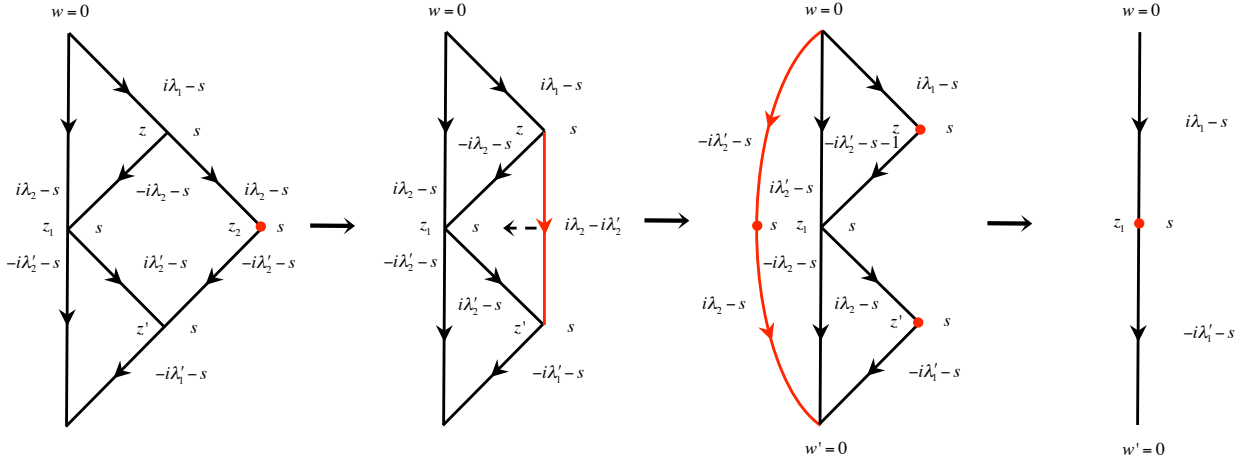}}
\end{picture}
}
\end{center}
\caption{\label{BosonicInnerReduction} Steps in evaluation of the bosonic inner product \re{bosonicproduct}.}
\end{figure}
%%%%%%%%%%%%%%%%%%%%%%%%%%%%%%%%%%%%%%%%%%%%%%%%%%%%%%%%%%%%%%%%%%%%%

For two excitations, the eigenfunctions of the matrix elements are given in Eqs.\ \re{2PlowestPyramid}, \re{2PhighestPyramid} for the same-flavor 
case and \re{2PmixedPyramid} for the mixed one. For $\Phi_{ss}$ and $\Phi_{s+1/2,s+1/2}$ eigenfunctions, the proof of the orthogonality condition 
repeats the steps of the bosonic consideration \cite{Belitsky:2014rba}. Namely, using a chain of transformations, exhibited in Fig.\ \ref{BosonicInnerReduction}, 
which consists of using (i) the chain rule \re{ChainRules}, (ii) the permutation identity \re{bosonicPermutationIdentity}, (iii) the chain rule (twice again), one reduces 
the inner product to the one-particle case, analyzed above, such that we immediately find
\begin{align}
\label{bosonicproduct}
\vev{\Phi_{ss} (\bit{\lambda}') | \Phi_{ss} (\bit{\lambda})}
=
a_{s} (s - i \lambda_1, s + i \lambda'_2) 
a_{s} (s + i \lambda'_1, s - i \lambda_2) 
\vev{\Phi_{s} (\lambda'_2) | \Phi_{s} (\lambda_2)}
\vev{\Phi_{s} (\lambda'_1) | \Phi_{s} (\lambda_1)}
\, .
\end{align}
Here, the inner product involves the spin-$s$ component in the Grassmann expansion of the one-particle eigenfunction \re{1PmatrixElement}.

To understand what to anticipate for the mixed $\Phi_{s,s+1/2}$ and $\Phi_{s+1/2,s}$ eigenfunctions, let us point out that we are dealing with
a degenerate case. Namely, the two-particle mixed sector can be cast in the following matrix form
\begin{align*}
\left(
\begin{array}{cc}
H_{11} & H_{12} \\
H_{21} & H_{22}
\end{array}
\right)
\left(
\begin{array}{c}
\ket{\psi_1} \\
\ket{\psi_2}
\end{array}
\right)
=
E
\left(
\begin{array}{c}
\ket{\psi_1} \\
\ket{\psi_2}
\end{array}
\right)
\, ,
\end{align*}
where the two eigenstates $\ket{\psi_1} \to \Phi_{21}$ and $\ket{\psi_2} \to \Phi_{12}$ share the same eigenvalue $E$, see Eq.\ \re{2PHeigenvalue}.
Then multiplying this equation from the left by the conjugate two-vector of eigenfunctions, we find
\begin{align*}
(E^\prime - E)
\left[
\vev{\psi^\prime_1|\psi_1}
+
\vev{\psi^\prime_2|\psi_2}
\right]
=
0
\, ,
\end{align*}
so that
\begin{align*}
\vev{\psi^\prime_1|\psi_1}
+
\vev{\psi^\prime_2|\psi_2}
=
\delta (E^\prime - E)
\, ,
\end{align*}
and not separately for each eigenstate. Thus, the orthogonality has to emerge from the sum of integrals
\begin{align}
\label{MixedInnerProductStart}
\vev{\Phi_{s,s+1/2} (\bit{\lambda}')|\Phi_{s,s+1/2} (\bit{\lambda})}
&
+
\vev{\Phi_{s+1/2,s} (\bit{\lambda}')|\Phi_{s+1/2,s} (\bit{\lambda})}
\\
&=
\int [D z_1]_{s +1/2} \int [D z_2]_{s} \Big( \Phi_{s+1/2,s} (\bit{z}; \bit{\lambda}') \Big)^\ast \Phi_{s+1/2,s} (\bit{z}; \bit{\lambda})
\nonumber\\
&+
\int [D z_1]_{s} \int [D z_2]_{s +1/2} \Big( \Phi_{s,s+1/2} (\bit{z}; \bit{\lambda}') \Big)^\ast \Phi_{s,s+1/2} (\bit{z}; \bit{\lambda})
\, . \nonumber
\end{align}
Making use of the pyramid representation for each eigenfunction
\begin{align}
\label{psi1Two}
\Phi_{s+1/2,s} (\bit{z})
&
=
(s + i \lambda_2)
z_1^{i \lambda_2 - s}
\int [D z]_{s + 1/2} (z_1 - z^\ast)^{- i \lambda_2 - s - 1} (z_2 - z^\ast)^{i \lambda_2 - s} z^{i \lambda_1 - s - 1/2}
\, , \\
\label{psi2Two}
\Phi_{s,s+1/2} (\bit{z})
&
=
(s - i \lambda_2)
z_1^{i \lambda_2 - s}
\int [D z]_{s + 1/2} (z_1 - z^\ast)^{- i \lambda_2 - s} (z_2 - z^\ast)^{i \lambda_2 - s - 1} z^{i \lambda_1 - s - 1/2}
\, ,
\end{align}
a simple-minded application of the rules used in the bosonic subsector fails at the second step. In spite of the fact that one can find a way out of
this predicament by using inversion\footnote{We would like to thank Sasha Manashov for this suggestion.} as demonstrated in Appendix 
\ref{2PorthogonalityAlternative}, we will follow a different route that can be applied for pentagon transitions studied later in the paper.

%%%%%%%%%%%%%%%%%%%%%%%%%%%%%%%%%%%%%%%%%%%%%%%%%%%%%%%%%%%%%%%%%%%%%
%            Figure
%%%%%%%%%%%%%%%%%%%%%%%%%%%%%%%%%%%%%%%%%%%%%%%%%%%%%%%%%%%%%%%%%%%%%
\begin{figure}[p]
\begin{center}
\mbox{
\begin{picture}(0,490)(250,0)
\put(0,-50){\insertfig{30}{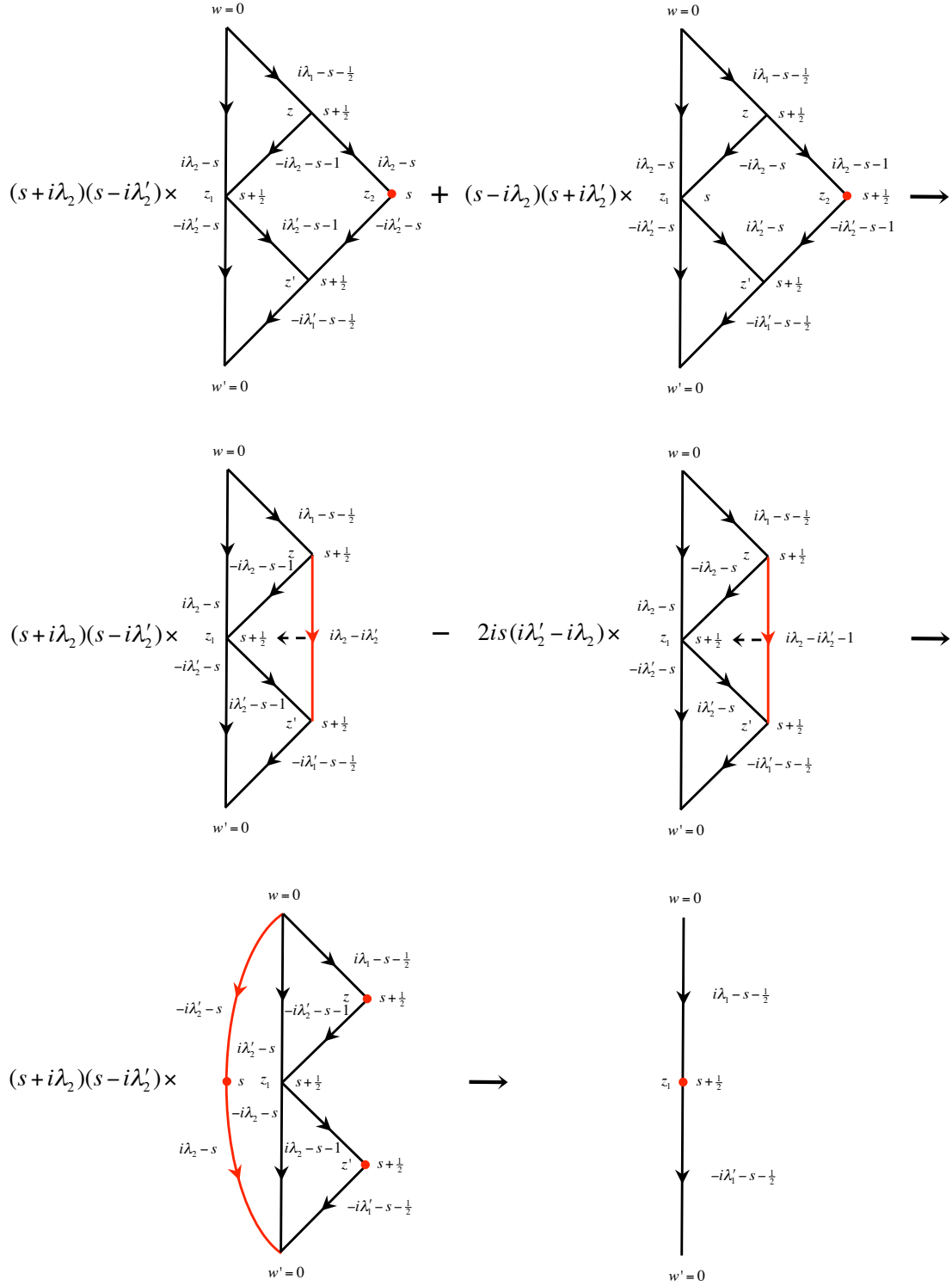}}
\end{picture}
}
\end{center}
\caption{\label{FigMixedInnerProduct} Reduction steps in the evaluation of the inner product \re{MixedInnerProductStart}.}
\end{figure}
%%%%%%%%%%%%%%%%%%%%%%%%%%%%%%%%%%%%%%%%%%%%%%%%%%%%%%%%%%%%%%%%%%%%%

First, we introduce a conjugate pyramid. It will be defined by the same graph as the original one but with all lines reversed and changed sign
of all rapidities. It is proportional to the complex conjugate wave function up to a phase factor that can be easily established from the involution rules 
\begin{align}
\big( (z' - z^\ast)^{-\alpha} \big)^\ast = {\rm e}^{i \pi \alpha^\ast}  (z - z'^\ast)^{-\alpha^\ast}
\, . 
\end{align}
This way the wave function $\left( \Phi_{s+1/2,s} (\bit{z}) \right)^\ast$ is determined by the reversed graph with overall phase factor
\begin{align}
\label{2Ptphase}
{\rm e}^{i \pi (s + i \lambda_2)} {\rm e}^{i \pi (s + 1/2 + i \lambda_1)} {\rm e}^{i \pi (2s + 1)}
\, .
\end{align}
Here, the first two factors stem from lines connecting vertices with $w = 0$ and the rest arise from the internal lines. The same prefactor accompanies 
the definition of $\left( \Phi_{s,s+1/2} (\bit{z}) \right)^\ast$. 

Now we proceed with the verification of orthogonality. The sum in Eq.\ \re{MixedInnerProductStart} is shown by the top row in Fig.\ \ref{FigMixedInnerProduct}, up to an 
overall phase \re{2Ptphase}, where all rapidities have to be dressed with primes since they emerge from the wave function in the out state. Starting the reduction from right 
to left, we integrate first with respect to the vertex $z_2$ by means of the chain rules \re{ChainRules}. This will yield different $a$-factors that accompany the reduced 
graphs, due to the different spins of the corresponding integration measures. Pulling out the overall factor
$$
{\rm e}^{- i \pi s} a_s (s - i \lambda_2, s + i \lambda'_2) 
$$
the two contributions with corresponding rapidity-dependent coefficients are shown in the middle row in Fig.\ \ref{FigMixedInnerProduct}. The subsequent reduction is 
based on the use of the permutation identity in superspace, which allows us to move the right vertical propagator through the entire graph to the left. To achieve this, we 
choose the coordinates as
\begin{align}
Z_1 = (w, 0)
\, , \qquad
Z_2 = (z, \theta)
\, , \qquad
Z'_1 = (w', 0)
\, , \qquad
Z'_2 = (z', \theta')
\, ,
\end{align}
in Eq.\ \re{superPermutationIdentity}, where obviously $w = w' = 0$. Collecting terms accompanying the Grassmann structure $\theta' \theta^\ast$, we find 
the relation
\begin{align}
\label{PermutationIdentityBosonFermion}
&
(z' - z^\ast)^{i \lambda' - i \lambda} (s - i \lambda') (s + i \lambda) X_{s + 1/2}  (w,  z; \lambda | w', \, z'; \lambda')
\nonumber\\
&\qquad
-
2 i s (i \lambda' - i \lambda) (z' - z^\ast)^{i \lambda' - i \lambda - 1} X_{s} (w, z, ; \lambda | w', z' ; \lambda')
\nonumber\\
&\qquad\qquad
=
(w' - w^\ast)^{i \lambda - i \lambda'} (s - i \lambda) (s + i \lambda') X_{s + 1/2}  (w, z ; \lambda' | w',  z';  \lambda)
\, ,
\end{align}
between the crosses with the spin-$s$ and spin-$(s+\ft12)$ measures,
\begin{align}
&
X_{s} (w, z; \lambda|w', \, z' ; \lambda')
\\
&\qquad
\equiv 
\int [D y]_s (y - w^\ast)^{i \lambda - s} (w' - y^\ast)^{- i \lambda' - s} (y - z^\ast)^{- i \lambda - s} (z' - y^\ast)^{i \lambda' - s}
\, , \nonumber\\
&
X_{s + 1/2}  (w, z; \lambda | w', z'; \lambda')
\\
&\qquad
\equiv 
\int [D y]_{s + 1/2} (y - w^\ast)^{i \lambda - s} (w' - y^\ast)^{- i \lambda' - s} (y - z^\ast)^{- i \lambda - s - 1} (z' - y^\ast)^{i \lambda' - s - 1}
\, . \nonumber
\end{align}
These arise from the Grassmann expansion of the supercross. We can recognize right away, in the left-hand side of Eq.\ \re{PermutationIdentityBosonFermion}, 
the sum of contributions with correct accompanying coefficients in Fig.\ \ref{FigMixedInnerProduct} (middle row). This allows us to use this permutation identity and pass 
to the leftmost graph in the bottom row of diagrams in Fig.\ \ref{FigMixedInnerProduct}, where we relied on the identity \re{OrthogIdentity} for $w' = w = 0$ which yielded the 
inner product of one-particle spin-$s$ component eigenfunctions \re{1PmatrixElement} along with the corresponding phase,
\begin{align}
(s - i \lambda_2) (s + i \lambda'_2)  {\rm e}^{- i \pi (s + i \lambda'_2)} \vev{\Phi_s (\lambda'_2)| \Phi_s (\lambda_2)}
\, .
\end{align}
We also included the overall factor of rapidities that stems from the right-hand side coefficient in the permutation identity \re{PermutationIdentityBosonFermion}.
This completes the first level in recursive reduction. 

At the next step, we use the chain rule twice, at the vertices $z$ and $z'$. This procedure generates the multiplicative factors
\begin{align}
{\rm e}^{- i \pi (2s + 1)} a_{s+1/2} (\ft12 + s - i \lambda_1, 1 + s + i \lambda'_2) a_{s+1/2} (\ft12 + s + i \lambda'_1, 1 + s - i \lambda_2) 
\, ,
\end{align}
accompanying the integral that can be computed by means of the chain rule (rightmost graph in the last row of Fig.\ \ref{FigMixedInnerProduct}), or rather the orthogonality 
identity, giving
\begin{align}
{\rm e}^{- i \pi (s + 1/2 + i \lambda'_1)} \vev{\Phi_{s + 1/2} (\lambda'_1) | \Phi_{s + 1/2} (\lambda_1)}
\, .
\end{align}
Combining everything together, we realize that all phases cancel out and we end up with the anticipated orthogonality relation
\begin{align}
\vev{\Phi_{s,s+1/2} (\bit{\lambda}')|\Phi_{s,s+1/2} (\bit{\lambda})}
&
+
\vev{\Phi_{s+1/2,s} (\bit{\lambda}')|\Phi_{s+1/2,s} (\bit{\lambda})}
\\
&=
(s - i \lambda_2) (s + i \lambda'_2)
a_{s+1/2} (\ft12 + s - i \lambda_1, 1 + s + i \lambda'_2) 
\nonumber\\
&\times
a_{s+1/2} (\ft12 + s + i \lambda'_1, 1 + s - i \lambda_2) 
\vev{\Phi_{s} (\lambda'_2) | \Phi_{s} (\lambda_2)}
\vev{\Phi_{s + 1/2} (\lambda'_1) | \Phi_{s + 1/2} (\lambda_1)}
, \nonumber
\end{align}
in terms of the individual one-particle component \re{1PmatrixElement} inner products defined in Eq.\ \re{Component1PinnerProduct}.

Since the procedure is inductive, the above reduction procedure suffices in the proof of the generic $N$-site case.

\section{From matrix elements to wave functions}

In the previous sections, we were dealing with the matrix elements \re{GenericMatrixElement} of the flux-tube operators that diagonalize the light-cone
Hamiltonian \re{TotalHamiltonian}. Let us pass to the flux-tube wave function $\bit{\Psi}_s  (\bit{X}; \bit{ \lambda})$ of $N$ excitations, --- 
localized at supercoordinates $\bit{X} = (X_1, X_2, \dots, X_N)$ where $X_n = (x_n, \vartheta_n)$ with the bosonic component belonging to the real axis, i.e., 
$\Im{\rm m} [x_n] = 0$, and having rapidities $\bit{\lambda} = (\lambda_1, \lambda_2, \dots, \lambda_N)$, --- that underlines the physics of the 
flux-tube for scattering amplitudes. A flux-tube state $\ket{E (\bit{\lambda}) }$ can be represented in its terms as
\begin{align}
\ket{E (\bit{\lambda}) }
 =
\int_{\mathcal{S}} d^N \bit{x} \int d^N \bit{\vartheta} \, \bit{\Psi}_s  (\bit{X}; \bit{\lambda})  \mathcal{O}_\Pi (\bit{X}) \ket{0}
\, ,
\end{align}
where the differenial measures are
\begin{align}
d^N \bit{x} = d x_1 d x_2 \dots dx_N
\, , \qquad
d^N \bit{\vartheta} = d \vartheta_1 d \vartheta_2 \dots d \vartheta_N
\end{align}
and the integration with respect to the bosonic variables is performed over the simplex $\mathcal{S} = \{ \infty > x_N \geq x_{N-1} \geq  \dots \geq x_1 \geq 0 \}$. Notice 
that bosonic and fermionic content of corresponding components jump places in the wave function compared to the superfield operator, e.g., for one-particle 
$\bit{\Psi}_s (X_1; \lambda_1) = \Psi_{s + 1/2} (x_1; \lambda_1) + \vartheta_1 \Psi_s (x_1; \lambda_1)$, where, for instance, for the $s = 1/2$ case, the lowest and 
highest components are fermion and boson, respectively, i.e., opposite to the matrix element \re{1PmatrixElement}.

One can easily deduce the representation of the sl$(2|1)$ generators on the space of the wave functions. Making use of 
\begin{align}
\mathcal{G} \ket{E (\bit{\lambda})}
=
\int_{\mathcal{S}} d^N \bit{x} \int d^N \bit{\vartheta} \, \bit{\Psi} (\bit{X}; \bit{\lambda}) \sum_{n = 1}^N G_n \mathcal{O}_\Pi (\bit{X}) \ket{0}
\, ,
\end{align}
where $\mathcal{G}$ acts on the Hilbert space of the flux-tube states and $G$ being its representation on flux-tube superfields 
$[\mathcal{G} , \mathcal{O}_\Pi (\bit{X})]  = \sum_{n = 1}^N G_n \mathcal{O}_\Pi (\bit{X})$, and integrating by parts, we find the
representation $\widehat{G}$ on wave functions
\begin{align}
\int_{\mathcal{S}} d^N \bit{x} \int d^N \bit{\vartheta}  \, \bit{\Psi}_s  (\bit{X}; \bit{\lambda}) \sum_{n = 1}^N G_n \mathcal{O}_\Pi (\bit{X}) \ket{0}
=
\int_{\mathcal{S}} d^N \bit{x} \int d^N \bit{\vartheta}  \, \left( \sum_{n = 1}^N \widehat{G}_n \bit{\Psi}_s  (\bit{X}; \bit{\lambda}) \right) \mathcal{O}_\Pi (\bit{X}) \ket{0}
\, .
\end{align}
Their explicit expressions read
\begin{align}
&
\widehat{S}^- 
= 
\partial_x
\, ,  
\qquad\qquad
\widehat{S}^+ 
= - x^2 \partial_x + x (2 s - 1) - x \vartheta \partial_\vartheta
\, , 
\qquad\qquad
\widehat{S}^0 
= - x \partial_x + s - \ft12 - \ft12 \vartheta \partial_\vartheta
\, , \nonumber\\
&
\widehat{B} 
= - \ft12 \vartheta \partial_\vartheta - s + \ft12
\, , \quad
\widehat{V}^- 
= 
\partial_\vartheta
\, ,  \quad
\widehat{W}^- 
= 
\vartheta \partial_x
\, , \quad
\widehat{V}^+ 
= 
x \partial_\vartheta
\, , \quad
\widehat{W}^+ 
= 
\vartheta \left(x \partial_x + 1 - 2 s \right)
\, .
\end{align}

\subsection{Wave function Hamiltonians}

In this section, we will derive Hamiltonians acting on the space of wave functions. To start with, it is instructive to recall the bosonic case but we defer this discussion
to Appendix \ref{SL2WFhamiltonianAppendix}, which the reader should consult first. Below, we proceed directly to the sl$(2|1)$ case and address the problem 
in two ways, first, by integration by parts and, then, using an intertwiner.

\subsubsection{Integration by parts}

Since the two-particle case contains all required elements, i.e., the bulk and boundary Hamiltonians, 
\begin{align}
\mathcal{H}_2 = \mathcal{H}_{01} + \mathcal{H}_{12}^+ + \mathcal{H}_{12}^- + \mathcal{H}_{1\infty}
\, ,
\end{align}
as alluded to in Section \ref{MEhamiltonianSection}, we will use it as a representative example. Following the same steps as above, we can calculate the Hamiltonian 
for the sl$(2|1)$ wave function. The latter enters the definition of the two-particle state
\begin{align*}
\ket{E (\bit{\lambda})}  = \int_{\mathcal{S}} d^2 \bit{x} \int d^2 \bit{\vartheta} \bit{\Psi}_s (\bit{X}; \bit{\lambda}) \mathcal{O}_\Pi (\bit{X}) \ket{0}
\, .
\end{align*}
Starting with the action of the Hamiltonian $\mathbb{H}$ on the Hilbert space of flux-tube excitations, 
\begin{align}
\label{IntegrationByPartsSL21}
\mathbb{H} \ket{E (\bit{\lambda})}
=
\int_{\mathcal{S}} d^2 \bit{x} \int d^2 \bit{\vartheta}\, \bit{\Psi}_s (\bit{X}; \bit{\lambda}) \, \mathcal{H}  \mathcal{O}_\Pi (\bit{X}) \ket{0}
=
\int_{\mathcal{S}} d^2 \bit{x} \int d^2 \bit{\vartheta} \, \left( \widehat{\mathcal{H}} \bit{\Psi}_s (\bit{X}; \bit{\lambda}) \right) \mathcal{O}_\Pi (\bit{X}) \ket{0}
\, ,
\end{align}
we immediately obtain its integral representation on the space of wave functions
\begin{align}
\label{TwoParticleWaveHamiltonian}
\widehat{\mathcal{H}}_2
=
\widehat{\mathcal{H}}_{01} + \widehat{\mathcal{H}}_{12}^+ + \widehat{\mathcal{H}}_{12}^- + \widehat{\mathcal{H}}_{2\infty}
\end{align}
with individual components
\begin{align}
\widehat{\mathcal{H}}_{01} \bit{\Psi}_s (\bit{X})
&
=
\int_1^{x_2/x_1} \frac{d \alpha}{\alpha - 1}  \left[ \alpha^{1 - 2 s} \bit{\Psi}_s (\alpha X_1, X_2) - \frac{1}{\alpha} \bit{\Psi}_s (X_1, X_2) \right]
\, , \\
\widehat{\mathcal{H}}_{2\infty} \bit{\Psi}_s (\bit{X})
&
=
\int_{x_1/x_2}^1 \frac{d \alpha}{1 - \alpha} \left[ \bit{\Psi}_s (X_1, \alpha x_2, \vartheta_2) - \bit{\Psi}_s (X_1, X_2) \right]
\, , \\
\widehat{\mathcal{H}}_{12}^+ \bit{\Psi}_s (\bit{X})
&
=
\int_1^\infty \frac{d \alpha}{\alpha - 1} 
\\
&\times
\left[
\left(
\frac{\alpha x_2 - x_1}{x_2 - x_1} 
\right)^{1 - 2s}
\bit{\Psi}_s \left( X_1, \alpha x_2, \frac{\alpha x_2 - x_1}{x_2 - x_1} \vartheta_2 - \frac{(\alpha - 1) x_2}{x_2 - x_1} \vartheta_1 \right)
-
\frac{1}{\alpha}
\bit{\Psi}_s ( X_1, X_2)
\right]
\, , \nonumber\\
\widehat{\mathcal{H}}_{12}^- \bit{\Psi}_s (\bit{X})
&
=
\int_0^1 \frac{d \alpha}{1 - \alpha}
\\
&\times
\left[
\left(
\frac{x_2 - \alpha x_1}{x_2 - x_1} 
\right)^{1 - 2s}
\bit{\Psi}_s \left( \alpha x_1, \frac{x_2 - \alpha x_1}{x_2 - x_1} \vartheta_1 - \frac{(1 - \alpha) x_1}{x_2 - x_1} \vartheta_2, X_2 \right)
-
\bit{\Psi}_s (X_1, X_2)
\right]
\, , \nonumber
\end{align}
where, for brevity, we did not display the dependence of $\bit{\Psi}_s$ on $\bit{\lambda}$.

\subsubsection{Intertwiner}

A generalization of the intertwiner to the sl$(2|1)$ case is relatively straightforward. Rather than being multiplication by a function as in the bosonic 
case, see \cite{Belitsky:2014rba} and Appendix \ref{SL2WFhamiltonianAppendix}, it becomes an operator in Grassmann variables. Namely, it admits 
the following form 
\begin{align}
\label{SL21intertwiner}
\mathcal{W}_N
=
(2s)^{- N}
\int d \vartheta'_1 d \vartheta'_2 \dots d \vartheta'_N
\Big( (x_1 + \vartheta'_1 \vartheta_1) (x_{21} + \vartheta'_{21} \vartheta_{21}) \dots (x_{N,N-1} + \vartheta'_{N,N-1} \vartheta_{N,N-1})  \Big)^{2 s}
\, ,
\end{align}
and induces the change from the matrix element to the wave function representations
\begin{align}
\label{SL21Intertwiner}
\widehat{\mathcal{H}}_N \, \mathcal{W}_N = \mathcal{W}_N \, \mathcal{H}_N
\, .
\end{align}
More specifically, we deduce the following relations between the two-particle bulk and boundary Hamiltonians
\begin{align}
&
\widehat{\mathcal{H}}^-_{12} \, \mathcal{W}_2 = \mathcal{W}_2 \, \mathcal{H}_{01}
\, , \qquad
\widehat{\mathcal{H}}^+_{12} \, \mathcal{W}_2 = \mathcal{W}_2 \, \mathcal{H}_{2\infty}
\, , \nonumber\\
&
\widehat{\mathcal{H}}_{01} \, \mathcal{W}_2 = \mathcal{W}_2 \, \mathcal{H}_{12}^-
\, , \qquad
\widehat{\mathcal{H}}_{2\infty} \, \mathcal{W}_2 = \mathcal{W}_2 \, \mathcal{H}_{12}^+
\, .
\end{align}
Though the individual components of the Hamiltonians transform differently under the operation of integration by parts and
by means of the intertwiner, the total sums are obviously the same.

\subsection{Wave functions}

We can adopt the above intertwiner in order to find the form of wave functions from the eigenfunctions of matrix elements via the relation
\begin{align}
\label{PsiPhiIntertwining}
\bit{\Psi}_s (\bit{X}; \bit{\lambda})= \mathcal{W}_N \bit{\Phi}_s (\bit{X}; \bit{\lambda})
\, .
\end{align}
However, as a cross check of the formalism that we developed here, it is instructive to solve for them explicitly diagonalizing
the generator of conserved changes $\widehat{D}_N$. Below, we will limit ourselves to the case of one- and two-particle excitations.

For a single excitation, as a solution to the eigenvalue equation
\begin{align}
\widehat{D}_1 (i w) \bit{\Psi}_s (X_1; \lambda_1) 
= 
(i w + i  \lambda_1 + s - \ft12)  \Psi_{s+1/2} (x_1; \lambda_1)
+ 
(i w + i  \lambda_1 + s)  \vartheta_1 \Psi_{s} (x_1; \lambda_1)
\, ,
\end{align}
we find for the component wave functions
\begin{align}
\Psi_s (x_1; \lambda_1) 
&
= 
x_1^{2s - 1} \Phi_s (x_1, \lambda)
=
x_1^{i \lambda_1 + s - 1/2}
\, , \\
\Psi_{s + 1/2} (x_1; \lambda_1) 
&
= 
(2s)^{-1} x_1^{2s} \Phi_{s + 1/2} (x_1, \lambda)
=
(2s)^{-1}
x_1^{i \lambda_1 + s - 1}
\, .
\end{align}
These expressions are in agreement with the intertwining relation \re{PsiPhiIntertwining} with the one-particle matrix element from 
Eq.\ \re{1PmatrixElement}.

Moving on to two excitations, the eigenfunction is decomposed in the component form as
\begin{align}
\bit{\Psi}_s (\bit{X}; \bit{\lambda}) 
= 
\Psi_{s+1/2, s+1/2} (\bit{x}) +  \vartheta_1 \Psi_{s, s + 1/2} (\bit{x})
+
\vartheta_2 \Psi_{s + 1/2, s} (\bit{x}) +  \vartheta_1  \vartheta_2 \Psi_{ss} (\bit{x})
\, ,
\end{align} 
and the solution to the eigenvalue equation for $\widehat{D}_2$
\begin{align}
\widehat{D}_2 (i w) \bit{\Psi}_s (\bit{X}; \bit{\lambda}) 
&
= 
(i w + i  \lambda_1 + s - \ft12)  (i w + i  \lambda_2 + s - \ft12)  \Psi_{s+1/2,s+1/2} (\bit{x}; \bit{\lambda})
\\
&
+ 
(i w + i  \lambda_1 + s )  (i w + i  \lambda_2 + s )  \vartheta_1 \vartheta_2 \Psi_{s,s} (\bit{x}; \bit{\lambda})
\nonumber\\
&
+
(i w + i  \lambda_1 + s - \ft12)  (i w + i  \lambda_2 + s)  
\left[
\vartheta_1 \Psi_{s,s+1/2} (\bit{x}; \bit{\lambda})
+
\vartheta_2 \Psi_{s+1/2,s} (\bit{x}; \bit{\lambda})
\right]
\, , \nonumber
\end{align}
generates the solutions
\begin{align}
\Psi_{s + 1/2, s + 1/2} (\bit{x}) 
& 
=
- (2s)^{-2}
x_1^{i \lambda_1 + s - 1/2} x_2^{i \lambda_2 + s - 1/2} \left( 1 - \frac{x_1}{x_2} \right)^{2s} {_2F_1}
\left.\left(
{ s + \ft12 + i \lambda_1,  s + \ft12 - i \lambda_2 \atop 2s + 1}
\right|
1 - \frac{x_1}{x_2}
\right)
\, , \\
\Psi_{ss} (\bit{z}) 
&
= 
x_1^{i \lambda_1 + s - 1} x_2^{i \lambda_2 + s - 1} \left( 1 - \frac{x_1}{x_2} \right)^{2s - 1} {_2F_1}
\left.\left(
{ s + i \lambda_1,  s - i \lambda_2 \atop 2s}
\right|
1 - \frac{x_1}{x_2}
\right)
\, , \\
\label{Solpsi1}
\Psi_{s, s + 1/2} (\bit{x}) 
&
= 
x_1^{i \lambda_1 + s - 1/2} x_2^{i \lambda_2 + s - 1} \left( 1 - \frac{x_1}{x_2} \right)^{2s - 1} {_2F_1}
\left.\left(
{ s + \ft12 + i \lambda_1,  s - i \lambda_2 \atop 2s}
\right|
1 - \frac{x_1}{x_2}
\right)
\, , \\
\label{Solpsi2}
\Psi_{s + 1/2, s} (\bit{x}) 
&
=
-
x_1^{i \lambda_2 + s} x_2^{i \lambda_1 + s - 3/2} \left( 1 - \frac{x_1}{x_2} \right)^{2s - 1} {_2F_1}
\left.\left(
{ s + \ft12 - i \lambda_1,  s + i \lambda_2 \atop 2s}
\right|
1 - \frac{x_1}{x_2}
\right)
\, .
\end{align}
A simple use of well-known connection formulas for hypergeometric functions, allows one to rewrite these expressions as a sum
of incoming and outgoing waves with modulated profiles. For $s = 1/2$, we reproduce results also obtained in Ref.\ \cite{BasSchVie15} found by diagonalizing
$\Omega_2$ defined in Eq.\ \re{OmegaN}. Acting with \re{TwoParticleWaveHamiltonian} on the wave function derived above, we find the expected eigenvalues
\begin{align}
\widehat{\mathcal{H}} \bit{\Psi}_s (\bit{X}; \bit{\lambda})
&
=
(E_{s + 1/2} (\lambda_1) + E_{s + 1/2} (\lambda_2)) \Psi_{s + 1/2, s + 1/2} (\bit{x}; \bit{\lambda}) 
+
\vartheta_1 \vartheta_2 (E_{s} (\lambda_1) + E_{s} (\lambda_2)) \Psi_{ss}  (\bit{x}; \bit{\lambda}) 
\nonumber\\
&
+
(E_{s + 1/2} (\lambda_1) + E_{s} (\lambda_2)) 
\left[
\vartheta_1 \Psi_{s, s + 1/2}  (\bit{x}; \bit{\lambda}) 
+
\vartheta_2 \Psi_{s + 1/2, s} (\bit{x}; \bit{\lambda}) 
\right]
\, . \end{align}
The intertwiner involves all components at a given order in Grassmann decomposition,
\begin{align}
\label{MixedIntertwiner}
\mathcal{W}_2 \Phi^{[g]}_{2} =  \Psi^{[g]}_{2}
\, ,
\end{align}
where $\Psi^{[g]}$ with $N \geq g \geq 0$ is defined in the same fashion as for the matrix element.
Using the two-particle $\mathcal{W}_2$, we can verify the above formulas by means of well-known relations between hypergeometric functions
\begin{align}
\label{2PbosonicWF}
\Psi_{ss} (\bit{x}; \bit{\lambda}) 
&
= 
(x_1 x_{21})^{2s - 1} \Phi_{ss} (\bit{x}; \bit{\lambda}) 
\, , \\
\Psi_{s+1/2, s+1/2} (\bit{x}; \bit{\lambda}) 
&
= 
- (2s)^{-2} (x_1 x_{21})^{2s} \Phi_{s+1/2, s+1/2} (\bit{x}; \bit{\lambda}) 
\, , \\
\label{2PfermionicWF12}
\Psi_{s, s+1/2}  (\bit{x}; \bit{\lambda}) 
&=
(2s)^{-1} (x_1 x_{21})^{2s - 1} 
\left[ x_1 \Phi_{s,s+1/2} + x_2 \Phi_{s+1/2, s} \right]  (\bit{x}; \bit{\lambda}) 
\, , \\
\label{2PfermionicWF21}
\Psi_{s+1/2,s}  (\bit{x}; \bit{\lambda}) 
&=
- 
(2s)^{-1} (x_1 x_{21})^{2s - 1} 
\left[ x_1 \Phi_{s,s+1/2} + x_1 \Phi_{s+1/2, s} \right]  (\bit{x}; \bit{\lambda}) 
\, .
\end{align}
These indeed coincide with Eqs.\ \re{PhiSS} -- \re{PhiS/2S/2}.

\section{Inner product on the line}

Making use of the above properties of the intertwining operator, we can introduce the following inner product for the boundary value of the matrix element eigenfunctions 
on the real line
\begin{align}
( \bit{\Phi}' | \bit{\Phi} )
\equiv
\int_{\mathcal{S}} d^N \bit{x} \int d^N \bit{\vartheta} \, \big( \bit{\Phi}' (\bit{X}) \big)^\ast \mathcal{W}_N \bit{\Phi} (\bit{X})
\, , 
\end{align}
where $\bit{X} = (X_1, \dots, X_N)$ with $X_n = (x_n, \vartheta_n)$. Employing Eqs.\ \re{SL21Intertwiner} 
and \re{IntegrationByPartsSL21}, it is straightforward to verify that the Hamiltonian is hermitian with respect to this inner product,
\begin{align}
( \bit{\Phi}' | \mathcal{H}_N \bit{\Phi} )
=
( \mathcal{H}_N \bit{\Phi}' | \bit{\Phi} )
\, .
\end{align}

We can relate the above inner product to the one in the upper half-plane of the complex plane. We substitute in form replying on the defining property of the reproducing
kernel
\begin{align}
\bit{\Phi} (\bit{X}) = \int [D^N \bit{Z}]_s \mathbb{K}_j (\bit{X}, \bit{Z}^\ast) \bit{\Phi} (\bit{Z})
\, , 
\end{align}
where $\bit{X} = (X_1, \dots, X_N)$ belongs to the real axis while $\bit{Z} = (Z_1, \dots Z_N)$ with $Z_n = (z_n, \theta_n)$ is being complex. The measure and reproducing kernels
are
\begin{align}
[D^N \bit{Z}]_s = \prod_{n = 1}^N [D Z_n]_s
\, , \qquad
\mathbb{K}_s (\bit{X}, \bit{Z}^\ast) =  \prod_{n = 1}^N \mathbb{K}_s (X_n, Z_n^\ast) 
\, , 
\end{align}
with
\begin{align}
\mathbb{K}_s (X, Z^\ast) = (x - z^\ast + \vartheta \theta^\ast)^{-2s}
\, .
\end{align}
Then we deduce the relation
\begin{align}
\label{RelationBetweenInnerProducts}
( \bit{\Phi}' | \bit{\Phi} ) = \vev{ \bit{\Phi}' | \mathcal{X}_N | \bit{\Phi}}
\, ,
\end{align}
where the operator $\mathcal{X}_N$ is determined by its integral kernel
\begin{align}
\mathcal{X}_N \bit{\Phi} (\bit{Z})
&
=
\int [D^N \bit{W}]_s \big( \mathbb{K}_s (\bit{Z}) |  \mathbb{K}_s (\bit{W}^\ast) \big) \bit{\Phi} (\bit{W})
\\
&
=
\int_{\mathcal{S}} d^N \bit{x} \int d^N \bit{\vartheta} \, 
\mathbb{K}_s (\bit{Z}, \bit{X}) \mathcal{W}_N \bit{\Phi} (\bit{X})
\, .
\end{align}
Finally, using the properties
\begin{align}
&
(S^0_Z + B_Z) \mathbb{K}_s (Z, X) = - (S^0_X - B_X) \mathbb{K}_s (Z, X)
\, , \qquad
S^\pm_Z \mathbb{K}_s (Z, X) = - S^\pm_X \mathbb{K}_s (Z, X)
\, , \\
&
W^+_Z \mathbb{K}_s (Z, X) = - V^+_X \mathbb{K}_s (Z, X)
\, , \qquad\qquad\qquad\qquad
V^-_Z \mathbb{K}_s (Z, X) = - W^-_X \mathbb{K}_s (Z, X)
\, ,
\end{align}
one can prove commutativity with $D_N$
\begin{align}
[\mathcal{X}_N, D_N] = 0
\, .
\end{align}

\subsection{Eigenvalues of $\mathcal{X}$}

Let us turn to evaluation of the eigenvalues of the operator $\mathcal{X}_N$ on the eigenfunctions $\Phi_s$.

\subsubsection{One excitation}

For the eigenfunction of one-particle matrix element, we find
\begin{align}
\mathcal{X}_1 \bit{\Phi}_s (Z_1; \lambda_1) = \int d x_1 \int d \vartheta_1 \mathbb{K}_j (Z_1, X_1) \mathcal{W}_1 \bit{\Phi}_s (X_1; \lambda_1)
\, , 
\end{align}
where, see Eq. \re{SL21intertwiner},
\begin{align}
\mathcal{W}_1 \bit{\Phi}_s (X_1) = \int d \vartheta'_1 (x_1 + \vartheta'_1 \vartheta_1) \bit{\Phi}_s (x_1, \vartheta'_1)
\, .
\end{align}
Substituting Eq.\ \re{1PmatrixElement}, we find 
\begin{align}
\mathcal{X}_1 \bit{\Phi}_s (Z_1; \lambda_1) 
=
\mathcal{X}_s (\lambda_1) \Phi_{s} (x_1; \lambda_1)
+
\mathcal{X}_{s + 1/2} (\lambda_1) \vartheta_1  \Phi_{s + 1/2} (x_1; \lambda_1)
\, ,
\end{align}
with the eigenvalues arising from the evaluation of the integral
\begin{align}
\label{1PXeigenvalue}
\mathcal{X}_s (\lambda_1) = \int_0^\infty d y (y - 1)^{- 2 s} y^{i \lambda_1 + s - 1}
=
{\rm e}^{- \pi (\lambda_1 + s)} \mu_s (\lambda_1)
\, ,
\end{align}
where the spin-$s$ flux-tube measure was introduced in Eq.\ \re{FTmeasureMUs}.

\subsubsection{Two excitations and more}
\label{TwoParticleIntegralsSection}

In the two-particle case, the action of $\mathcal{X}_2$ reads explicitly 
\begin{align}
\mathcal{X}_2 \bit{\Phi}_s (\bit{Z}; \bit{\lambda})
&
=
\int_{\mathcal{S}} d^2 \bit{x} \int d^2 \bit{\vartheta}
(z_1 - x_1 + \theta_1 \vartheta_1)^{-2s} (z_2 - x_2 + \theta_2 \vartheta_2)^{-2s}
\nonumber\\
&
\times
\int d^2 \bit{\vartheta}'
(x_1 + \vartheta'_1 \vartheta_1)^{2s} (x_{21} + \vartheta'_{21} \vartheta_{21})^{2s} \bit{\Phi}_s (x_1, \vartheta'_1, x_2, \vartheta'_2; \bit{\lambda})
\, .
\end{align}
For the lowest and highest Grassmann components, i.e., 
\begin{align}
\Phi^{[0]}_2 (\bit{X}; \bit{\lambda})
= \Phi_{ss} (\bit{X}; \bit{\lambda})
\, , \qquad
\Phi^{[2]}_2 (\bit{X}; \bit{\lambda})
=
\vartheta_1 \vartheta_2 \Phi_{s+1/2, s+1/2} (\bit{X}; \bit{\lambda})
\, ,
\end{align}
according to the terminology of Section \ref{AlgebraicEigenfunctionsSection}, we get the anticipated result as in the purely bosonic model \cite{Belitsky:2014rba} 
for different values of the conformal spin,
\begin{align}
\mathcal{X}_2 \Phi^{[0]}_2 (\bit{Z}; \bit{\lambda}) = \mathcal{X}_s (\lambda_1)  \mathcal{X}_s (\lambda_2) \Phi^{[0]}_2 (\bit{Z}; \bit{\lambda})
\, , \qquad
\mathcal{X}_2 \Phi^{[2]}_2 (\bit{Z}; \bit{\lambda}) = \mathcal{X}_{s + 1/2} (\lambda_1)  \mathcal{X}_{s + 1/2} (\lambda_2) \Phi^{[2]}_2 (\bit{Z}; \bit{\lambda})
\, .
\end{align}
These results can be easily found going to the asymptotic region $x_2 \gg x_1$ and making use of the asymptotic form of the eigenfunctions
$\Phi_{ss} (\bit{x}; \bit{\lambda}) \simeq x_1^{i \lambda_1 - s} x_2^{i \lambda_2 - s}$ and the same for $\Phi_{s+1/2, s+1/2}$ with an obvious shift of the spin.

For the mixed components, 
\begin{align}
\Phi^{[1]}_2 (\bit{X}; \bit{\lambda}) = \vartheta_1 \Phi_{s+1/2, s} (\bit{x}; \bit{\lambda}) + \vartheta_2 \Phi_{s, s+1/2} (\bit{x}; \bit{\lambda})
\, ,
\end{align}
the situation is trickier and we want to perform the diagonalization exactly. Namely, after performing the Grassmann integration we obtain
\begin{align}
\mathcal{X}_2 \Phi^{[1]}_2 (\bit{Z}; \bit{\lambda}) 
&=
\theta_1 \int_{\mathcal{S}} d^2 \bit{x} (z_1 - x_1)^{- 2 s - 1}(z_2 - x_2)^{- 2s} \Psi_{s+1/2, s} (\bit{x}; \bit{\lambda}) 
\\
&
+
\theta_2 \int_{\mathcal{S}} d^2 \bit{x} (z_1 - x_1)^{- 2s}(z_2 - x_2)^{- 2s - 1} \Psi_{s, s+1/2} (\bit{x}; \bit{\lambda}) 
\, ,
\end{align}
where the integrand is given in terms of two-particle wave functions \re{2PfermionicWF12} and \re{2PfermionicWF21}. A calculation, following the steps outlined in Appendix C.2 
of Ref.\ \cite{Belitsky:2014rba}, demonstrates that
\begin{align}
\label{IntPsi2Topsi1}
 \int_{\mathcal{S}} d^2 \bit{x} (z_1 - x_1)^{- 2s - 1}(z_2 - x_2)^{- 2s} \Psi_{s+1/2,s} (x_1, x_2) 
=
\mathcal{X}_{s+1/2, s} (\bit{\lambda}) \Phi_{s+1/2,s} (z_1, z_2) 
\, ,
\end{align}
and 
\begin{align}
\label{IntPsi1Topsi2}
\int_{\mathcal{S}} d^2 \bit{x} (z_1 - x_1)^{- 2s}(z_2 - x_2)^{- 2s-1} \Psi_{s,s+1/2} (x_1, x_2)
=
\mathcal{X}_{s+1/2, s} (\bit{\lambda}) \Phi_{s,s+1/2} (z_1, z_2) 
\, ,
\end{align}
so that $\Phi^{[1]}_2 (\bit{Z}; \bit{\lambda})$ is an eigenfunction of $\mathcal{X}_2$
\begin{align}
\label{MixedXeigenvalue}
\mathcal{X}_2 \Phi^{[1]}_2 (\bit{Z}; \bit{\lambda})
=
\mathcal{X}_{s+1/2, s} (\bit{\lambda}) \Phi^{[1]}_2 (\bit{Z}; \bit{\lambda})
\, ,
\end{align}
with the same eigenvalue for its both components
\begin{align}
\mathcal{X}_{s+1/2, s} (\bit{\lambda}) = \mathcal{X}_{s + 1/2} (\lambda_1) \mathcal{X}_s (\lambda_2) 
\, .
\end{align}
It is expressed in terms of the one-particle eigenvalues \re{1PXeigenvalue}.

This result immediately generalizes to any $N$. For the Grasssmann degree-$g$ $N$-particle case, we have
\begin{align}
\mathcal{X}_N \Phi^{[g]}_N (\bit{Z}; \bit{\lambda})
=
\left( \prod_{n = 1}^g \mathcal{X}_{s + 1/2} (\lambda_n) \right) \left( \prod_{n = g + 1}^N \mathcal{X}_{s} (\lambda_n) \right)
\Phi^{[g]}_N (\bit{Z}; \bit{\lambda})
\, ,
\end{align}
again observing factorization of multiparticle eigenvalues.

\section{Square transitions}
\label{SectionBoxTranstitions}

The $N$-particle wave functions have to be orthogonal with respect to the so-called square transitions, i.e., when both the incoming and outgoing states
are in the same conformal frame. Namely, we define
\begin{align}
\bit{B} (\bit{\lambda}| \bit{\lambda}' )
&
\equiv
\vev{E (\bit{\lambda}') | E (\bit{\lambda}) }
\nonumber\\
&
=
\int d^N \bit{X}' \int d^N \bit{X} \, \big( \bit{\Psi}_s (\bit{X}'; \bit{\lambda}') \big)^\ast \bit{G} (\bit{X}', \bit{X}) \bit{\Psi}_s (\bit{X}; \bit{\lambda})
\, ,
\end{align}
where $\bit{G}$ is a product 
\begin{align}
\bit{G} (\bit{X}', \bit{X}) = \prod_{n=1}^N G (X'_n, X_n) 
\end{align}
of supersymmetric propagators on the real axis
\begin{align}
G (X'_n, X_n) = \left( x_n + x'_n + \theta_n \theta'_n \right)^{- 2 s}
\, .
\end{align}

We will demonstrate its relation to the inner product  in the upper half of the complex plane of the matrix element eigenfunctions for the two-to-two transition. 
As before, since components of different Grassmann degree do not talk to each other, we would like to keep track of these by using a marker variable as in 
Section \ref{OrthogonalitySection}. So the decomposition of the supersquare transition into three independent Grassmann components is
\begin{align}
\bit{B} (\bit{\lambda} |\bit{\lambda}')
=
B_{s+1/2,s+1/2|s+1/2,s+1/2}  (\bit{\lambda} |\bit{\lambda}' )
+
\varepsilon \varepsilon'
B_{s,s+1/2|s,s+1/2}  (\bit{\lambda} | \bit{\lambda}' )
+
(\varepsilon \varepsilon')^2
B_{ss|ss}  (\bit{\lambda} | \bit{\lambda}' )
\, . 
\end{align}
These are related via the equations
\begin{align}
B_{s+1/2,s+1/2|s+1/2,s+1/2} = B_{s+1/2,s+1/2} \, , \quad  B_{s,s+1/2|s,s+1/2} = B_{s,s+1/2} +  B_{s+1/2,s}  \, , \quad  B_{ss|ss} = B_{ss}
\, ,  
\end{align}
to individual integrals $B_{s_1 s_2}$ involving spin-$(s_1, s_2)$ wave functions connected with propagators from top to bottom of the square 
\begin{align}
B_{s_1s_2} (\bit{\lambda} | \bit{\lambda}')
=
\int_{\mathcal{S}'} d^2 \bit{x}'  \int_{\mathcal{S}} d^2 \bit{x} 
\, 
\frac{\bit( \Psi_{s_1 s_2} (\bit{x}'; \bit{\lambda}') \big)^\ast \Psi_{s_1 s_2} (\bit{x}; \bit{\lambda}) }{(x'_1 + x_1)^{2 s_1} (x'_2 + x_2)^{2 s_2}}
\, .
\end{align}
These integrals were computed in Section \ref{TwoParticleIntegralsSection}, with the result
\begin{align}
\int_{\mathcal{S}} d^2 \bit{x}\, \Psi_{s_1 s_2} (\bit{x}; \bit{\lambda}) (x'_1 + x_1)^{- 2s_1} (x'_2 + x_2)^{-2s_2} =  \mathcal{X}_{s_1 s_2} (\bit{\lambda}) \Phi_{s_1 s_2} (\bit{x}'; \bit{\lambda})
\, , 
\end{align}
where $\mathcal{X}_{s_1 s_2} (\bit{\lambda})$ is the eigenvalue of the operator $\mathcal{X}_2$, such that
\begin{align}
B_{s_1 s_2}  (\bit{\lambda} | \bit{\lambda}')
\equiv
(\Phi_{s_1s_2} (\bit{\lambda}') | \Phi_{s_1s_2} (\bit{\lambda}))
=
\mathcal{X}_{s_1 s_2} (\bit{\lambda}) 
\int_{\mathcal{S}'} d^2 \bit{x}' \big( \Psi_{s_1 s_2} (\bit{x}'; \bit{\lambda}') \big)^\ast \Phi_{s_1s_2} (\bit{x}'; \bit{\lambda})
\, .
\end{align}
Since the wave functions $\Psi$ of the top and bottom components are related to the matrix elements $\Phi$ by means of a multiplicative factor of bosonic coordinates, we
recognize in the right-hand side of the above relation, the inner product in the upper half plane for $\Phi$. Their orthogonality was demonstrated in Ref.\ \cite{Belitsky:2014rba}, 
as well as was recapitulated above in Section \ref{SectionOrthogonalityPhi}. Only the mixed component require special attention. Starting from the relation between the inner 
products \re{RelationBetweenInnerProducts}, we can extract the mixed components and find
\begin{align}
&
\int_{\mathcal{S}} d^2 \bit{x}
\big[ \left( \Phi_{s+1/2, s} (\bit{x}; \bit{\lambda}') \right)^\ast \Psi_{s+1/2, s} (\bit{x}; \bit{\lambda})  
+ \left( \Phi_{s, s+1/2} (\bit{x}; \bit{\lambda}') \right)^\ast \Psi_{s, s+1/2} (\bit{x}; \bit{\lambda}')
\big]
\nonumber\\
&\qquad
=
\int_{\mathcal{S}} d^2 \bit{x}
\big[ \left( \Psi_{s+1/2, s} (\bit{x}; \bit{\lambda}') \right)^\ast \Phi_{s+1/2, s} (\bit{x}; \bit{\lambda})  + \left( \Psi_{s, s+1/2} (\bit{x}; \bit{\lambda}') \right)^\ast \Phi_{s, s+1/2} (\bit{x}; \bit{\lambda}')
\big]
\nonumber\\
&\qquad\qquad
=
\mathcal{X}_{s+1/2, s} (\bit{\lambda})
\left\{
\vev{ \Phi_{s+1/2, s} (\bit{\lambda}') |  \Phi_{s+1/2, s} (\bit{\lambda})} + \vev{ \Phi_{s,s+1/2} (\bit{\lambda}') |  \Phi_{s,s+1/2} (\bit{\lambda})} 
\right\}
\, .
\end{align}
In the right-hand side of the above equation, we used the eigenvalue equation \re{MixedXeigenvalue}, while in the left-hand side, we employed the relation between the matrix elements 
and wave functions \re{2PfermionicWF12} and \re{2PfermionicWF21}. The right-hand side of the above equation was calculated in Section \ref{SectionOrthogonalityPhi} and shows
orthogonality of wave functions with respect to the square transitions.

Generalization to $N$-particle square transitions goes along the same lines making use of results of the previous section.

\section{Pentagon transitions}
\label{SectionPentagonTranstitions}

The $N$-particle super-wave functions define the pentagon transitions, i.e., the building blocks of the Operator Product Expansion for scattering amplitudes as was reviewed in the
Introduction. Namely
\begin{align}
\bit{P} (\bit{\lambda} | \bit{\lambda}' )
&
\equiv
\vev{E (\bit{\lambda}') | \mathcal{P} | E (\bit{\lambda}) }
\nonumber\\
&
=
\int_{\mathcal{S}'} d^N \bit{X}' \int_{\mathcal{S}} d^N \bit{X} \, \big( \bit{\Psi}'_s (\bit{X}'; \bit{\lambda'}) \big)^{\ast} \bit{G} (\bit{X}', \bit{X}) \bit{\Psi}_s (\bit{X}; \bit{\lambda})
\, ,
\end{align}
where compared to the just discussed box transitions, the wave function in the final state is in a different conformal frame (to be specified later on) compared to the initial one. 
The reduction of the $N$-particle pentagon to $N-1$ pentagon goes through the same chain of transformation as the inductive proof for the orthogonality condition. Thus we will 
demonstrate it for first non-trivial case, i.e., two-site wave functions.

In complete analogy with the above consideration, one finds the following component expansion for the two-to-two pentagon transition
\begin{align}
\bit{P} (\bit{\lambda} | \bit{\lambda}' )
=
P_{s+1/2,s+1/2|s+1/2,s+1/2} (\bit{\lambda} | \bit{\lambda}' )
+
\varepsilon \varepsilon'
P_{s,s+1/2|s,s+1/2}   (\bit{\lambda} | \bit{\lambda}' )
+
(\varepsilon \varepsilon')^2
P_{ss|ss}  (\bit{\lambda} | \bit{\lambda}' )
\, , 
\end{align}
where we adopted a notation $P_{s_1s_2|s'_1,s'_2}$ used in the pentagon approach \cite{Basso:2013aha,Belitsky:2014rba,Basso:2014koa,Belitsky:2014sla,%
Basso:2014nra,Belitsky:2014lta,Belitsky:2015efa,Basso:2014hfa,Basso:2014jfa,Basso:2015rta,Fioravanti:2015dma,Belitsky:2015qla,Bonini:2015lfr,Belitsky:2015lzw} 
for particles with spins $(s_1,s_2)$ undergoing a transition to particles with spins  $(s'_1,s'_2)$. These are related via the equations
\begin{align}
P_{s+1/2,s+1/2|s+1/2,s+1/2} = P_{s+1/2,s+1/2} \, , \quad  P_{s+1/2,s|s+1/2,s} = P_{s,s+1/2} +  P_{s+1/s,s}  \, , \quad  P_{ss|ss} = P_{ss}
\, ,
\end{align}
to integrals involving an overlap of wave functions in different conformal frames
\begin{align}
P_{s_1,s_2} (\bit{\lambda}|\bit{\lambda}')
&
=
\int_{\mathcal{S}'} d^2 \bit{x}'  \int_{\mathcal{S}} d^2 \bit{x} \frac{\left( \Psi_{s_1s_2} (\bit{x}'; \bit{\lambda}') \right)^\ast \Psi_{s_1s_2} (\bit{x}; \bit{\lambda})}{(x_1 + x'_1)^{2 s_1} (x_2 + x'_2)^{2s_2}}
\, .
\end{align}
Before we move on to its calculation, we will take a detour by calculating the inverse wave functions first.

\subsection{Inversion of wave functions}

As we will show in the next subsection, the pentagon transitions can be reduced to the inner product of matrix elements inverted with
respect to the origin and one of them shifted away from it. Thus, we will introduce the operation of inversion
\begin{align}
z \to z^I = 1/z
\, ,
\end{align}
and construct the resulting wave functions. To start with, the spin-$s$ measure changes according to the rule
\begin{align}
[D z]_{s}
\to
[D z]^I_{s} =  (z z^\ast)^{- 2 s} [D z]_{s}
\, .
\end{align}
Since the same-flavor wave functions and corresponding pentagons were already discussed in Ref.\ \cite{Belitsky:2014rba}, we will not repeat it here. Thus,
we address only the mixed-flavor case. For $\Phi_{s+1/2, s}$ in Eq.\ \re{psi1Two}, we have
\begin{align}
\Phi^I_{s + 1/2, s} (\bit{z}; \bit{\lambda})
&
\equiv 
{\rm e}^{i \pi (2s + 1)}
z_1^{- 2 s - 1} z_2^{- 2s} \, \Phi_{s + 1/2, s} (\bit{z}^I; \bit{\lambda})
\nonumber\\
&=
(s + i \lambda_2)
\int [D z]_{s + 1/2} \, z^{- i \lambda_1 - s - 1/2} (z_1 - z^\ast)^{- i \lambda_2 - s - 1} (z_2 - z^\ast)^{i \lambda_2 - s}
z_2^{- i \lambda_2 - s}
\, ,
\end{align}
where $z_1^{- 2 s - 1} z_2^{- 2s}$ is the scaling factor with exponents proportional to the conformal weights of the points $z_1$ and $z_2$ and overall phase factor
was introduced to get rid of the one emerging from the inversion. Similarly we find for \re{psi2Two},
\begin{align}
\Phi^I_{s,s+1/2} (\bit{z}; \bit{\lambda})
&
\equiv 
{\rm e}^{i \pi (2s + 1)}
z_1^{- 2 s} z_2^{- 2s - 1} \, \Phi_{s,s + 1/2} (\bit{z}^I; \bit{\lambda})
\nonumber\\
&=
(s - i \lambda_2)
\int [D z]_{s + 1/2} z^{- i \lambda_1 - s - 1/2} (z_1 - z^\ast)^{- i \lambda_2 - s} (z_2 - z^\ast)^{i \lambda_2 - s - 1} z_2^{- i \lambda_2 - s}
\, .
\end{align}
The graphical representation for $\Phi^I_{s+1/2,s} $ is shown in Fig.\ \ref{FigPyramid} (on the right panel).

\subsection{Mixed pentagons}

Let us calculate the pentagon transitions corresponding to the wave function $\Psi_{s+1/2,s}$ and $\Psi_{s,s+1/2}$. They are
\begin{align}
P_{s,s+1/2} 
&
=
\int_{\mathcal{S}'} d^2 \bit{x}'  \int_{\mathcal{S}} d^2 \bit{x} \frac{\left( \Psi_{s,s+1/2} (\bit{x}'; \bit{\lambda}') \right)^\ast \Psi_{s,s+1/2} (\bit{x}; \bit{\lambda})}{(x_1 + x'_1)^{2 s} (x_2 + x'_2)^{2s+1}}
\, , \\
P_{s+1/2,s} 
&
=
\int_{\mathcal{S}'} d^2 \bit{x}'  \int_{\mathcal{S}} d^2 \bit{x} \frac{\left( \Psi_{s+1/2,s} (\bit{x}'; \bit{\lambda}') \right)^\ast \Psi_{s+1/2,s} (\bit{x}; \bit{\lambda})}{(x_1 + x'_1)^{2 s + 1} (x_2 + x'_2)^{2s}}
\, ,
\end{align}
where the wave functions are connected point-by-point with spin-$s$ propagators. They are related to the matrix elements via Eqs.\ \re{2PfermionicWF12} and 
\re{2PfermionicWF21}. Notice that the wave function in the out-state is in a different conformal frame with respect to the incoming ones, i.e., 
\begin{align}
\Psi'_{s_1s_2} (\bit{x}'; \bit{\lambda})  = \left( \frac{\partial x''_1}{\partial x'_1} \right)^{1 - s_1} \left( \frac{\partial x''_2}{\partial x'_2} \right)^{1 -s_2} \Psi_{s_1 s_2}  
(\bit{x}''; \bit{\lambda}) 
\, , \qquad
x'' = \frac{x'}{1 - x'}
\, .
\end{align}

Making use of the integrals displayed in Eqs.\ \re{IntPsi2Topsi1} and \re{IntPsi1Topsi2}, we can rewrite the above pentagons as
\begin{align}
P_{s,s+1/2} (\bit{\lambda}'|\bit{\lambda}) 
&
=
\mathcal{X}_{s+1/s,2} (\bit{\lambda})
\int_{\mathcal{S}'} d^2 \bit{x}' \left( \Psi'_{s,s+1/2} (\bit{x}';\bit{\lambda}') \right)^\ast \Phi_{s,s+1/2} (\bit{x};\bit{\lambda}) 
\, , \\
P_{s+1/2,s} (\bit{\lambda}'|\bit{\lambda}) 
&
=
\mathcal{X}_{s+1/s,2} (\bit{\lambda})
\int_{\mathcal{S}'} d^2 \bit{x}' \left( \Psi'_{s+1/2,s} (\bit{x}';\bit{\lambda}') \right)^\ast \Phi_{s+1/2,s} (\bit{x};\bit{\lambda}) 
\, .
\end{align}
It is important to realize that individually we cannot relate these integrals to the inner product on the line since the intertwiner $\mathcal{W}$ (see Eq.\ \re{SL21intertwiner})
acts on the superwave function $\Phi_2^{[2]} = \theta_1 \Phi_{s+1/2,s} + \theta_2 \Phi_{s, s+1/2}$, not its separate components $\Phi_{s_1 s_2}$ as shown in Eq.\ \re{MixedIntertwiner}.

Further, to relate the product on the line to the one in upper half-plane, one has to take into account that the operator $\mathcal{X}$ has well-defined eigenvalue again
only on the total mixed superfunction \re{MixedXeigenvalue}, not on its components in the Grassmann decomposition. This immediately implies that (as shown in Section 
\ref{SectionBoxTranstitions}) 
\begin{align}
&
\int_{\mathcal{S}} d^2 \bit{x}
\Big[ 
\left( \Psi'_{s+1/2,s} (\bit{x}; \bit{\lambda}') \right)^\ast \Phi_{s+1/2,s} (\bit{x}; \bit{\lambda}) 
+
\left( \Psi'_{s,s+1/2} (\bit{x}; \bit{\lambda}') \right)^\ast \Phi_{s,s+1/2} (\bit{x}; \bit{\lambda}) 
\Big]
\nonumber\\
&\qquad
=
\mathcal{X}_{s+1/2,s}
\Big[ 
\vev{ 
\Phi'_{s+1/2,s} (\bit{x}; \bit{\lambda}')
|
\Phi_{s+1/2,s} (\bit{x}; \bit{\lambda})
}
+
\vev{ 
\Phi'_{s,s+1/2} (\bit{x}; \bit{\lambda}')
|
\Phi_{s,s+1/2} (\bit{x}; \bit{\lambda})
}
\Big]
\, .
\end{align}
Therefore, we can relate the mixed pentagon to the sum of the inner products of the matrix element eigenfunctions in the upper half-plane of the complex plane, i.e., 
\begin{align}
P_{s+1/2,s}  (\bit{\lambda}' |\bit{\lambda}')
+
P_{s,s+1/2}  (\bit{\lambda}' |\bit{\lambda}')
=
\vev{ \Phi'_{s+1/2,s} (\bit{\lambda}') |  \Phi_{s+1/2,s} (\bit{\lambda})}  + \vev{ \Phi'_{s,s+1/2} (\bit{\lambda}') |  \Phi_{s,s+1/2} (\bit{\lambda})} 
\, .
\end{align}
The  latter can be rewritten as
\begin{align}
\label{PentagonsInnerProducts}
\vev{ \Phi'_{s+1/2,s} (\bit{\lambda}') |  \Phi_{s+1/2,s} (\bit{\lambda})} 
&
+ \vev{ \Phi'_{s,s+1/2'} (\bit{\lambda}') |  \Phi_{s,s+1/2} (\bit{\lambda})} 
\\
&
=
\vev{ \Phi^I_{s+1/2,s} (\bit{\lambda}'; 0) |  \Phi^I_{s+1/2,s} (\bit{\lambda}; 1)} 
+ 
\vev{ \Phi^I_{s,s+1/2} (\bit{\lambda}'; 0) |  \Phi^I_{s,s+1/2} (\bit{\lambda}; 1)}
\, , \nonumber
\end{align}
in terms of the inverted eigenfunctions at position $- \gamma$,
\begin{align}
\Phi^I_{s+1/2,s} (\bit{\lambda}; \gamma)
&
=
(s + i \lambda_2)
\nonumber\\
&\times
\int [D z]_{s + 1/2} \, (z + \gamma)^{- i \lambda_1 - s - 1/2} (z_1 - z^\ast)^{- i \lambda_2 - s - 1} (z_2 - z^\ast)^{i \lambda_2 - s}
(z_2 + \gamma)^{- i \lambda_2 - s}
\, , \\
\Phi^I_{s,s+1/2} (\bit{\lambda}; \gamma)
&=
(s - i \lambda_2)
\nonumber\\
&\times
\int [D z]_{s + 1/2} (z + \gamma)^{- i \lambda_1 - s - 1/2} (z_1 - z^\ast)^{- i \lambda_2 - s} (z_2 - z^\ast)^{i \lambda_2 - s - 1} (z_2 + \gamma)^{- i \lambda_2 - s}
\, .
\end{align}

\begin{figure}[p]
\begin{center}
\mbox{
\begin{picture}(0,490)(260,0)
\put(0,-50){\insertfig{30}{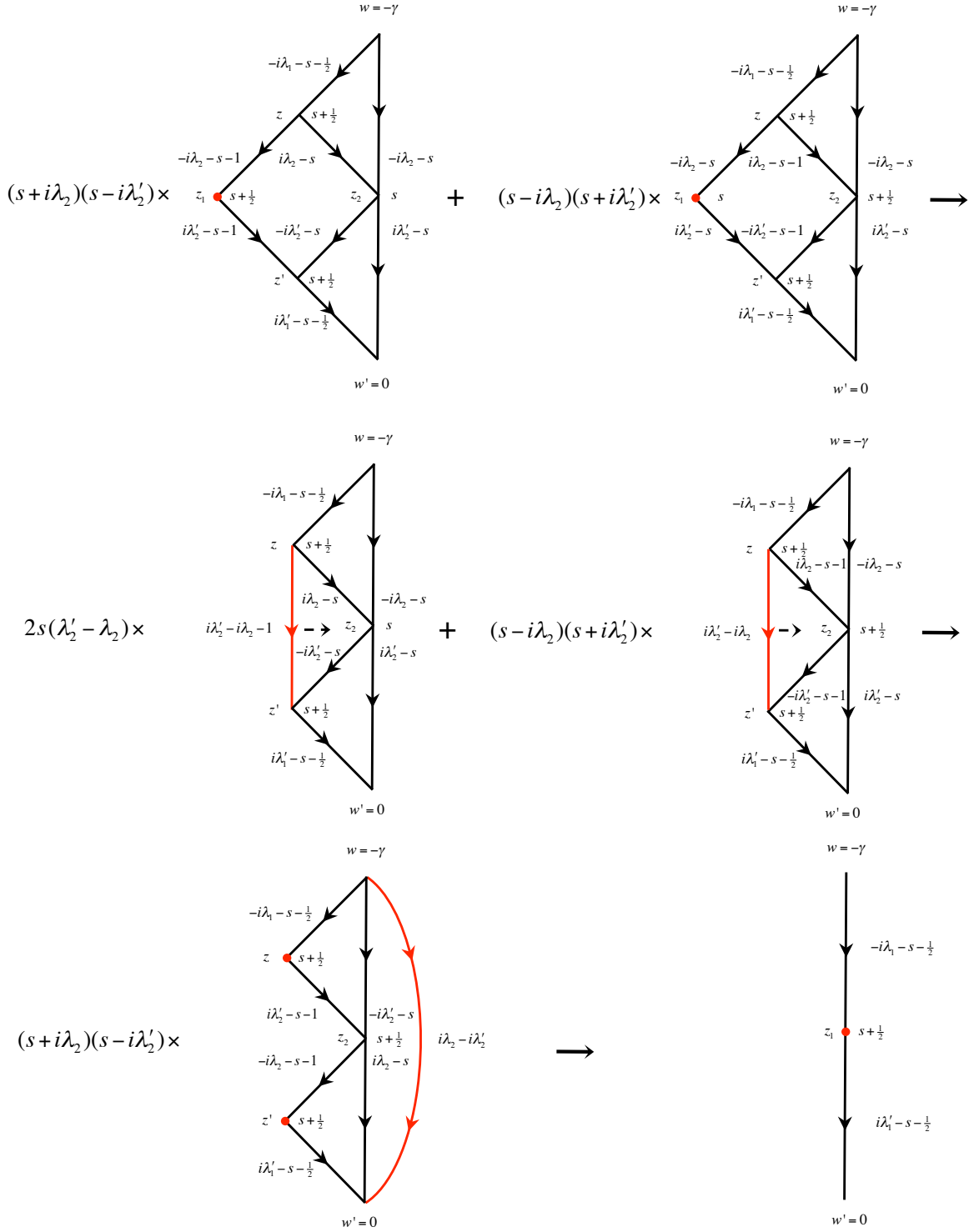}}
\end{picture}
}
\end{center}
\caption{\label{FigPentagon} Procedure for the calculation of the mixed pentagons \re{PentagonsInnerProducts}.}
\end{figure}

Let us calculate the transition
\begin{align}
T^\gamma_{s+1/2,s} (\bit{\lambda}'|\bit{\lambda})
=
\vev{ \Phi^I_{s+1/2,s} (\bit{\lambda}'; 0) |  \Phi^I_{s+1/2,s} (\bit{\lambda}; \gamma)} 
+ 
\vev{ \Phi^I_{s,s+1/2} (\bit{\lambda}'; 0) |  \Phi^I_{s,s+1/2} (\bit{\lambda}; \gamma)}
\, , 
\end{align}
for a generic values of $\gamma$. The calculation of the inner product in the right-hand side of this equation goes along the same lines as the one for the evaluation of
the inner product discussed in Section \ref{SectionOrthogonalityPhi}, this time, only from left to right. The first step, shown in the first line of Fig.\ \ref{FigPentagon}, consists 
in the integration over the vertex at $z_1$ making use of the chain rule \re{ChainRules} and pulling out the overall factor
$$
{\rm e}^{- i \pi s} a_s (s + i \lambda_2, s - i \lambda'_2)
$$
from both contributions, which yields the relative coefficient for the first graph
\begin{align}
{\rm e}^{- i \pi/2} \frac{(s + i \lambda_2)(s - i \lambda'_2) a_{s+1/2} (1 + s + i \lambda_2, 1 + s - i \lambda'_2)}{a_s (s + i \lambda_2, s - i \lambda'_2)}
=
2 s (\lambda_2 - \lambda'_2)
\, .
\end{align}
as shown in Fig.\ \ref{FigPentagon} (middle row diagrams). 

%%%%%%%%%%%%%%%%%%%%%%%%%%%%%%%%%%%%%%%%%%%%%%%%%%%%%%%%%%%%%%%%%%%%%
%            Figure
%%%%%%%%%%%%%%%%%%%%%%%%%%%%%%%%%%%%%%%%%%%%%%%%%%%%%%%%%%%%%%%%%%%%%
%%%%%%%%%%%%%%%%%%%%%%%%%%%%%%%%%%%%%%%%%%%%%%%%%%%%%%%%%%%%%%%%%%%%%

The subsequent step requires the application of the superpermutation identity. For the case at hand, we introduce the following supercoordinates 
in Eq.\ \re{superPermutationIdentity} to do the job
\begin{align}
Z_1 = (z, \theta)
\, , \qquad
Z_2 = (- \gamma, 0)
\, , \qquad
Z'_1 = (z', \theta')
\, , \qquad
Z'_2 = (0, 0)
\, ,
\end{align}
and rapidities $\lambda' = \lambda'_2$, $\lambda = \lambda_2$. Keeping the $\theta' \theta^\ast$ term in the Grassmann expansion on both sides 
of the permutation identity, we find the relation
\begin{align}
\label{SuperCrossGrassmannComponent}
&
(z' - z^\ast)^{i \lambda'_2 - i \lambda_2} (s - i \lambda_2) (s + i \lambda'_2) X_{s + 1/2}  (z, - \gamma; \lambda_2 | z', 0; \lambda'_2)
\nonumber\\
&\qquad
-
2 i s (i \lambda'_2 - i \lambda_2) (z' - z^\ast)^{i \lambda'_2 - i \lambda_2 - 1} X_{s} (z, - \gamma ; \lambda_2 | z', 0 ; \lambda'_2)
\nonumber\\
&\qquad\qquad
=
\gamma^{i \lambda_2 - i \lambda'_2} (s - i \lambda'_2) (s + i \lambda_2) X_{s + 1/2}  (z, - \gamma; \lambda'_2 | z', 0; \lambda_2)
\, ,
\end{align}
between crosses with spin-$s$ and spin-$(s+\ft12)$ measures,
\begin{align}
X_{s} (z, \gamma; \lambda_2 | z', 0; \lambda'_2 )
&
=
\int [D y]_{s}  (y - z^\ast)^{i \lambda_2 - s}  (y + \gamma)^{-i \lambda_2 - s} (z' - y^\ast)^{- i \lambda'_2 - s} (- y^\ast)^{i \lambda'_2 - s}
\, , \\
X_{s + 1/2} (z, \gamma; \lambda_2 | z', 0; \lambda'_2)
&
=
\int [D y]_{s + 1/2} (y - z^\ast)^{i \lambda_2 - s - 1}  (y + \gamma)^{-i \lambda_2 - s - 1} (z' - y^\ast)^{- i \lambda'_2 - s} (- y^\ast)^{i \lambda'_2 - s}
\, .
\end{align}
Finally, using the chain rule three times, we acquire factors
\begin{align}
{\rm e}^{- i \pi (2 s + 1)} a_{s+1/2} (\ft12 + s + i \lambda_1, 1 + s - i \lambda'_2) a_{s+1/2} (\ft12 + s - i \lambda'_1, 1 + s + i \lambda_2)
\end{align}
and 
\begin{align}
{\rm e}^{- i \pi (s + 1/2)} a_{s+1/2} (\ft12 + s + i \lambda_1, \ft12 + s - i \lambda'_1)
\, ,
\end{align}
from the steps shown in Fig.\ \ref{FigPentagon} in the left and right panels of the last row graphs, respectively.

Assembling everything together (along with phases that emerge from conjugated pyramid), we find
\begin{align}
T_{s+1/2,s}^\gamma (\bit{\lambda} | \bit{\lambda}')
&
=
{\rm e}^{- \pi \sum_{n = 1}^2 \lambda_n} \gamma^{i \sum_{n = 1}^2 (\lambda_n - \lambda'_n) }
(s + i \lambda_2) (s - i \lambda'_2)
a_{s+1/2} (\ft12 + s + i \lambda_1, \ft12 + s - i \lambda'_1)
\nonumber\\
&\times
a_s (s + i \lambda_2, s - i \lambda'_2)
a_{s+1/2} (\ft12 + s + i \lambda_1, 1 + s - i \lambda'_2) a_{s+1/2} (\ft12 + s - i \lambda'_1, 1 + s + i \lambda_2)
\, .
\end{align}
This relation implies factorizable structure of multiparticle pentagons, i.e., two-to-two in the current case, 
\begin{align}
P_{s+1/2,s| s+1/2,s} (\bit{\lambda} | \bit{\lambda}')
&
=
{\rm e}^{- \pi \sum_{n = 1}^2 \lambda_n}
\\
&\times
P_{s+1/2|s+1/2} (\lambda_1 | \lambda'_1)
P_{s+1/2|s} (\lambda_1 | \lambda'_2)
P_{s|s+1/2} (\lambda_2 | \lambda'_1)
P_{s|s} (\lambda_2 | \lambda'_2)
\, , \nonumber
\end{align}
in terms of one-particle pentagon transitions
\begin{align}
P_{s|s}  (\lambda | \lambda') 
&
= \frac{\Gamma (i \lambda - i \lambda') \Gamma (2s)}{\Gamma (s + i \lambda) \Gamma (s - i \lambda')}
\, , \\
P_{s+1/2|s+1/2}  (\lambda | \lambda') 
&
= \frac{\Gamma (i \lambda - i \lambda') \Gamma (2s + 1)}{\Gamma (s + \ft12 + i \lambda) \Gamma (s + \ft12 - i \lambda')}
\, , \\
P_{s|s + 1/2}  (\lambda | \lambda') 
&
= \frac{\Gamma (\ft12 + i \lambda - i \lambda') \Gamma (2 s+ 1)}{\Gamma (s + i \lambda) \Gamma (s + \ft12 - i \lambda')}
\, .
\end{align}
Finally $P_{s+1/2|s}  (\lambda | \lambda') = P_{s|s + 1/2}  (- \lambda' | - \lambda)$. Had we chosen the normalization of the wave functions $\Psi_{s,s+1/2}$
and $\Psi_{s+1/2,s}$ according to the coordinate Bethe Anzats, such that the asymptotic incoming wave come with a unit amplitude, we would cancel the 
prefactor ${\rm e}^{- \pi \sum_{n = 1}^2 \lambda_n}$ in the right-hand side of the above relation as well as recover pentagons $P_{s+1/2|s} (\lambda'_1|\lambda'_2)
P_{s|s+1/2} (\lambda_2|\lambda_1)$ entering the denominator of the transitional factorized form of the pentagons, as was shown for bosonic case in 
Ref.\ \cite{Belitsky:2014rba}. We will not do it here though, since the above form already proves the factorized form of multiparticle pentagons 
\cite{Basso:2013vsa,Belitsky:2015efa}. Generalization to arbitrary $N$ is straightforward since the procedure is inductive.

\section{Conclusions}

In this paper we solved an open superspin chain model that describes minimally supersymmetric sectors of the $\mathcal{N}=4$ flux tube.
Depending on the conformal spin assignment for the lowest component of the supermultiplet of flux-tube fields, it encodes either
hole-fermion or fermion-gluon excitations. The bulk interactions between the adjacent superfields building up the light-cone operators 
inherit the sl$(2|1)$ invariance of four-dimensional theory, however, the presence of the boundary breaks it down to the diagonal
subgroup. Using the factorized R-matrix structure of the sl$(2|1)$ symmetric spin chain, we constructed the eigenfunctions of the model, 
analytically continued to the upper half of the complex plane, in the form of multiple integrals which admit an intuitive Feynman graph
representation. The latter was indispensable for analytical proof of their orthogonality. The same framework was applied to calculate 
the so-called pentagon transitions between the states of the flux-tube in different conformal planes. The latter serve as building blocks
in the framework of the Operator Product Expansion to null polygonal Wilson superloops. The outcome of this analysis revealed 
factorizable structure of the dynamical part of multiparticle pentagons in terms of single-particle ones as was already extensively used 
in the past.

Our consideration can be extended to include all propagating modes of the maximally supersymmetric Yang-Mills theory and encode
them in the noncompact sl$(2|4)$ superchain. From the technical point of view the changes would appear to be minimal: one would have to replace
single Grassmann variables $\theta$ by an SU(4) vector $\theta^A$ and any product of conjugate ones by the sum, $\theta \theta^\ast
\to \theta^A \theta^\ast_A$. The question remains open however whether this construction will be able to encode and unravel the matrix 
part of the pentagon transitions.

\section*{Acknowledgments}

I would like to thank Benjamin Basso, Fidel Schaposhnik and Pedro Viera for rekindling my interest in the problem and participating
at the initial stage of the work. I am deeply indebted to Sasha Manashov for multiple insightful and clarifying discussions at later stages 
of the project and Didina Serban for useful conversations. This research was supported by the U.S. National Science Foundation under 
the grants PHY-1068286 and PHY-1403891.

\appendix

\section{Feynman graph primer}
\label{FeynmanGraphsAppendix}

In this appendix we provide a reminder of Feynman graphs used in the representation of eigenfunctions and some elementary operations on
them. The main building block in the construction is the propagator that receives Schwinger parametrization valid in the upper half of the
complex plane 
\begin{align}
\label{SchwingerParametrization}
(z' - z^\ast)^{- \alpha} = \frac{{\rm e}^{- i \pi \alpha/2}}{\Gamma (\alpha)} \int_0^\infty d p \, p^{\alpha - 1} {\rm e}^{i p (z' - z^\ast)}
\end{align}
and which is instrumental in all explicit calculations. Whenever the propagator's exponent $\alpha$ coincides with (twice) the spin of the integration 
measure, it becomes the so-called reproducing kernel that obeys the following defining property
\begin{align}
\int [Dz]_s (w - z^\ast)^{- 2s}  \Phi (z) = {\rm e}^{- i \pi s} \Phi (w)
\, .
\end{align}

%%%%%%%%%%%%%%%%%%%%%%%%%%%%%%%%%%%%%%%%%%%%%%%%%%%%%%%%%%%%%%%%%%%%%
%            Figure
%%%%%%%%%%%%%%%%%%%%%%%%%%%%%%%%%%%%%%%%%%%%%%%%%%%%%%%%%%%%%%%%%%%%%
\begin{figure}[t]
\begin{center}
\mbox{
\begin{picture}(0,90)(290,0)
\put(0,-500){\insertfig{28}{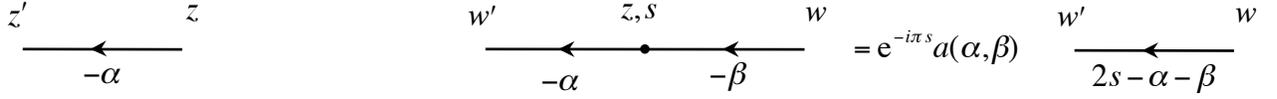}}
\end{picture}
}
\end{center}
\caption{\label{FigPropagator} Graphical representation of the propagator \re{SchwingerParametrization} (left panel) and the chain rule \re{ChainRules} (right panel).}
\end{figure}
%%%%%%%%%%%%%%%%%%%%%%%%%%%%%%%%%%%%%%%%%%%%%%%%%%%%%%%%%%%%%%%%%%%%%

The propagators obey the following set of rules useful in proofs of various relations.

\vspace{3mm}

\noindent $\bullet$ Chain rule:
\begin{align}
\label{ChainRules}
\int [Dz]_s (w' - z^\ast)^{- \alpha} (z - w^\ast)^\beta = {\rm e}^{- i \pi s} a_s (\alpha, \beta) (w' - w^\ast)^{2s - \alpha - \beta}
\, ,
\end{align}
where
\begin{align}
a_s (\alpha, \beta) = \frac{\Gamma (2s - \alpha - \beta) \Gamma (2s)}{\Gamma (\alpha) \Gamma (\beta)}
\, .
\end{align}
It is shown in Fig.\ \ref{FigPropagator}.

\vspace{3mm}

\noindent $\bullet$ Orthogonality identity:
\begin{align}
\label{OrthogIdentity}
a_s (s - i \lambda, s + i \lambda') (w' - w^\ast)^{i \lambda -  i \lambda'} |_{w = w' = 0}
=
\frac{2 \pi \Gamma (2s)}{\Gamma (s - i \lambda) \Gamma (s + i \lambda')} \delta (\lambda - \lambda')
\, .
\end{align}
This is easily verified by applying the above chain rules from right to left and Eq.\ \re{SchwingerParametrization}.

\vspace{3mm}

\noindent $\bullet$ Permutation identity:
\begin{align}
\label{bosonicPermutationIdentity}
(z'_1 - z_1^\ast)^{i \lambda' - i \lambda} X \left( \bit{z}; \lambda | \bit{z}'; \lambda' \right)
=
X \left( \bit{z}; \lambda' | \bit{z}'; \lambda \right) (z'_2 - z_2^\ast)^{i \lambda - i \lambda'}
\, ,
\end{align}
with the bosonic cross
\begin{align}
X (\bit{z}, \lambda | \bit{z}', \lambda')
\equiv
\int [D w]_s (w - z_1^\ast)^{i \lambda - s} (w - z_2^\ast)^{- i \lambda - s} (z'_1 - w^\ast)^{- i \lambda' - s} (z'_2 - w^\ast)^{i \lambda' - s}
\, ,
\end{align}
where $\bit{z} = (z_1, z_2)$ and $\bit{z}' = (z'_1, z'_2)$. The proof for this relation can be found in Appendix B of \cite{Belitsky:2014rba} and references cited therein.

\section{Alternative proof of orthogonality}
\label{2PorthogonalityAlternative}

%%%%%%%%%%%%%%%%%%%%%%%%%%%%%%%%%%%%%%%%%%%%%%%%%%%%%%%%%%%%%%%%%%%%%
%            Figure
%%%%%%%%%%%%%%%%%%%%%%%%%%%%%%%%%%%%%%%%%%%%%%%%%%%%%%%%%%%%%%%%%%%%%
\begin{figure}[t]
\begin{center}
\mbox{
\begin{picture}(0,180)(295,0)
\put(0,-410){\insertfig{28}{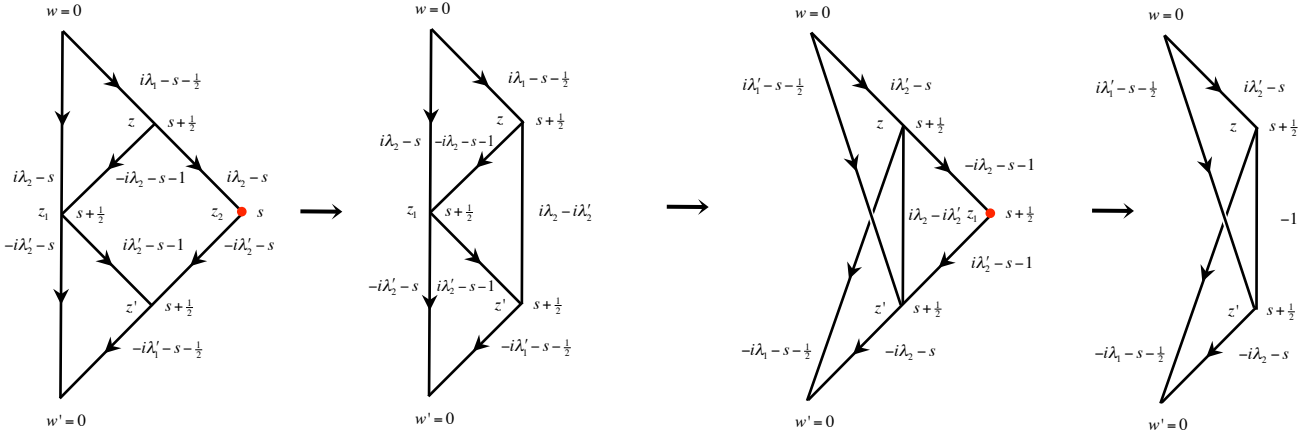}}
\end{picture}
}
\end{center}
\caption{\label{FigAlternativeInner} Proof of the orthogonality of the mixed inner product \re{OrthoMix} using inversion.}
\end{figure}
%%%%%%%%%%%%%%%%%%%%%%%%%%%%%%%%%%%%%%%%%%%%%%%%%%%%%%%%%%%%%%%%%%%%%

As we pointed out in the main text, instead of using the permutation identity for superpropagators, one can prove the orthogonality of mixed eigenfunctions
using a more `down-to-earth' procedure. Though, it is not generalizable to pentagon transitions, we think it is worthwhile to demonstrate it by showing
that
\begin{align}
\label{OrthoMix}
\left[
\vev{\Phi_{s+1/2,s} (\bit{\lambda}')|\Phi_{s+1/2,s} (\bit{\lambda})} 
+ 
\vev{\Phi_{s,s+1/2} (\bit{\lambda}')|\Phi_{s,s+1/2} (\bit{\lambda})} 
\right]_{\bit{\scriptstyle \lambda}' \neq \bit{\scriptstyle \lambda}} 
= 
0
\, .
\end{align}

We start with $\Phi_{s+1/2,s}$ and perform the integration with respect to $z_2$. This yields according to the chain rule \re{ChainRules} the propagator 
connecting the points $z'$ and $z$ with the exponent $i \lambda_2 - i \lambda'_2$ multiplied by the factor of rapidities $a_s (s - i \lambda_2, s + i \lambda'_2)$,
as shown in Fig.\ \ref{FigAlternativeInner} (second panel). Next we perform the variable transformation $z \to 1/z$ that transforms the graph into the one in Fig.\ 
\ref{FigAlternativeInner} (third panel). The subsequent integration with respect to the vertex $z_1$ becomes trivial via the chain rule and we end up with the 
integral $\mathcal{I}$
\begin{align}
\mathcal{I}
&
\equiv
\int [Dz]_{s + 1/2}
\int [Dz']_{s + 1/2}
(z' - z^\ast)^{-1}
z^{i \lambda'_2 - s} (z')^{i \lambda'_1 - s - 1/2} (- z^\ast)^{- i \lambda_1 - s - 1/2} (- z'^\ast)^{- i \lambda'_2 - s}
\, . \nonumber
\end{align}
represented by the rightmost graph in Fig.\ \ref{FigAlternativeInner} along with the factor $a_{s + 1/2}  (s + 1 + i \lambda_2, s + 1 - i \lambda'_2)$. So combining 
everything together, we obtain (up to an inessential overall phase)
\begin{align}
\label{Con1}
\vev{\Phi_{s+1/2,s} | \Phi_{s+1/2,s}}
=
(\ft12 + i \lambda_2)(\ft12 - i \lambda'_2) a_{s} (s - i \lambda_2, s + i \lambda'_2) a_{s + 1/2}  (s + 1 + i \lambda_2, s + 1 - i \lambda'_2) \, 
\mathcal{I}
\, .
\end{align}
Analogous reduction applies to the matrix element of $\Phi_{s,s+1/2}$ and we find (up to the same phase factor)
\begin{align}
\label{Con2}
\vev{\Phi_{s,s+1/2} | \Phi_{s,s+1/2}}
=
(\ft12 - i \lambda_2)(\ft12 + i \lambda'_2) a_{s} (s + i \lambda_2, s - i \lambda'_2) a_{s + 1/2}  (s + 1 - i \lambda_2, s + 1 + i \lambda'_2) \, 
\mathcal{I}
\, ,
\end{align}
with the very same integral $\mathcal{I}$! Summing up the two contributions \re{Con1} and \re{Con2} together one observes that the sum of the coefficients accompanying 
$\mathcal{I}$ cancels out between the two, so we verify the validity of \re{OrthoMix} even without knowing the explicit form\footnote{Which can be easily calculated 
in fact using Feynman parametrization for the propagator $(z' - z^\ast)^{-1}$ and one of the adjacent factors involving $z'$ to $z^\ast$ and subsequent use of
the chain rule. This demonstrates that $\mathcal{I} \sim \delta (\lambda_1 + \lambda_2 - \lambda'_1 - \lambda'_2)$.} of $\mathcal{I}$.

\section{Wave-function Hamiltonian in sl(2) sector}
\label{SL2WFhamiltonianAppendix}

In this Appendix, we recall the derivation of the Hamiltonians on the space of wave functions starting with the ones acting on matrix elements.
We start by performing this transformation for the one-particle state and then move on to two particles which involves all bulk and boundary interactions
intrinsic to an $N$-particle case.

The one-particle flux-tube state with energy $E (\lambda)$ is given by
\begin{align}
\ket{E (\lambda)} = \int_0^\infty dx_1 \, \psi_s (x_1; \lambda) O_\Pi (x_1) \ket{0}
\, ,
\end{align}
where the field $O_\Pi (x_1) = W^\dagger (0) \phi_s (x_1) W (\infty)$ creates an excitation out of the vacuum with $\psi_s (x_1; \lambda)$ being the eigenfunction 
of the Hamiltonian with the eigenvalue $E (\lambda)$. Our goal here, knowing the from of $H$ acting on $O_\Pi$, find $\widehat{H}$ for $\psi_s$.
The action of $\mathbb{H}$ on $\ket{E (\lambda)}$ is
\begin{align}
\label{HonBosonicState}
\mathbb{H} \ket{E (\lambda)} = \int_0^\infty dx_1 \, \psi_s (x_1; \lambda) (H_{01} + H_{1\infty}) O_\Pi (x_1) \ket{0}
\, ,
\end{align}
with
\begin{align}
H_{01} O_\Pi (x_1)
&
=
\int_0^1 \frac{d \beta}{1 - \beta} \left[ \beta^{2s - 1} O_\Pi (\beta x_1)  - O_\Pi (x_1) \right]
\, , \\
H_{1\infty} O_\Pi (x_1)
&=
\int_1^\infty \frac{d \beta}{\beta - 1} \left[O_\Pi (\beta x_1)  - \beta^{-1} O_\Pi (x_1) \right] 
\, .
\end{align}
Changing the integration variable in the right-hand side of Eq.\ \re{HonBosonicState} as $x_1 \to x_1/\beta$ and $\beta = 1/\alpha$, we immediately find
\begin{align}
\int_0^\infty dx_1 \, \psi_s (x_1; \lambda) H O_\Pi (x_1) = \int_0^\infty dx_1 \left( \widehat{H} \psi_s (x_1; \lambda) \right) O_\Pi (x_1) 
\, ,
\end{align}
with
\begin{align}
\widehat{H}_{01} \psi_s (x_1)
&
=
\int_1^\infty \frac{d \alpha}{\alpha - 1} \left[  \alpha^{1 - 2 s} \psi_s (\alpha x_1) - \alpha^{-1} \psi_s (x_1) \right]
\, , \\ 
\widehat{H}_{1\infty}  \psi_s (x_1)
&
=
\int_0^1 \frac{d \alpha}{1 - \alpha} \left[  \psi_s (\alpha x_1) -  \psi_s (x_1) \right]
\, ,
\end{align}
where, for brevity, we dropped the dependence of the wave function on the rapidity $\lambda$.

Next we turn to the case of two flux-tube excitations with rapidities $\bit{\lambda} = (\lambda_1, \lambda_2)$ that will unravel two questions: (i) how the strong ordering 
of coordinates in multiparticle wave functions enters the game as we as (ii) the form of the bulk Hamiltonians. As we will see from our calculation, the bulk and boundary
Hamiltonians jump places as we pass from matrix elements to wave functions. The state is
\begin{align}
\ket{E (\bit{\lambda})} 
= 
\int_0^\infty d^2 \bit{x} \, \theta (x_2 - x_1) \psi_s (\bit{x}; \bit{\lambda}) O_\Pi (\bit{x}) \ket{0}
\, ,
\end{align}
where $O_\Pi (\bit{x}) = W^\dagger (0) \phi_s (x_1) \phi_s (x_2) W (\infty)$ creates two particles localized at $\bit{x} = (x_1, x_2)$ integrated with the measure $d^2 \bit{x} = 
dx_1 \, dx_2$. The Hamiltonian splits up into the bulk $H^\pm_{12}$ and boundary  $H_{01/1\infty}$ terms,
\begin{align}
\mathbb{H} \ket{E (\bit{\lambda)}} = \int_0^\infty d^2 \bit{x} \, \theta (x_2 - x_1) \psi_s (\bit{x}; \bit{\lambda}) (H_{01} + H_{12}^+ + H_{12}^-+ H_{1\infty}) O_\Pi (\bit{x}) \ket{0}
\, ,
\end{align}
with
\begin{align}
H_{01} O_\Pi (\bit{x}) 
&
=
\int_0^1 \frac{d \beta}{1 - \beta} \left[ \beta^{2 s - 1}  O_\Pi (\beta x_1, x_2) - O_\Pi (x_1, x_2) \right]
\, , \\
H^+_{12} O_\Pi (\bit{x}) 
&= 
\int_{x_1/x_2}^1 \frac{d \beta}{1 - \beta}
\left[
\left( \frac{\beta x_2 - x_1}{x_2 - x_1} \right)^{2 s - 1}  O_\Pi (x_1, \beta x_2)
-
O_\Pi (x_1, x_2) 
\right]
\, , \\
H^-_{12} O_\Pi (\bit{x}) 
&= 
\int_1^{x_2/x_1} \frac{d \beta}{\beta - 1}
\left[
\left( \frac{x_2 - \beta x_1}{x_2 - x_1} \right)^{2 s - 1} O_\Pi (\beta x_1, x_2)
-
\beta^{-1} O_\Pi (x_1, x_2)  
\right]
\, , \\
H_{2 \infty} O_\Pi (\bit{x}) 
&
=
\int_0^1 \frac{d \beta}{1 - \beta} \left[ O_\Pi (x_1, \beta x_2) - \beta^{-1} O_\Pi (x_1, x_2) \right]
\, .
\end{align}
Again, the Hamiltonian on the space of wave functions is found by means of the integration by parts
\begin{align}
\int_0^\infty d^2 \bit{x} \, \theta (x_2 - x_1) \psi_s (\bit{x}; \bit{\lambda}) H O_\Pi (\bit{x}) 
=
\int_0^\infty d^2 \bit{x} \, \theta (x_2 - x_1) \left( \widehat{H} \psi_s (\bit{x}; \bit{\lambda}) \right) O_\Pi (\bit{x}) 
\, ,
\end{align}
with particular attention paid to strong ordering. In the $H_{01}$ term, we change the integration variable as $x_1 \to x_1/\beta$ and immediately find that the step-function imposes 
a lower limit on the range of $\beta$, $\beta > x_1/x_2$. Thus, substituting $\beta = 1/\alpha$, we find
\begin{align}
\widehat{H}_{01} \psi_s (\bit{x})
=
\int_1^{x_2/x_1} \frac{d \alpha}{\alpha - 1} \left[ \alpha^{1 - 2 s} \psi_s (\alpha x_1, x_2) - \alpha^{-1} \psi_s (x_1, x_2) \right]
\, .
\end{align}
Similarly for $H_{2\infty}$, we substitute $x_2 \to x_2/\beta$ and get the constraint on $\beta$, $x_2/x_1 > \beta$. Inverting $\beta$, $\beta = 1/\alpha$, we eventually obtain
\begin{align}
\widehat{H}_{2\infty} \psi_s (\bit{x})
=
\int_{x_1/x_2}^1 \frac{d \alpha}{1 - \alpha} \left[ \psi_s (x_1, \alpha x_2) - \psi_s (x_1, x_2) \right]
\, .
\end{align}
As we can see, the boundary Hamiltonians takes on the form of the bulk one. In an analogous manner, we derive $\widehat{H}^\pm_{12}$ following the same route as above
\begin{align}
\widehat{H}^+_{12} \psi_s (\bit{x})
&
=
\int_1^\infty \frac{d \alpha}{\alpha - 1}
\left[
\left( \frac{\alpha x_2 - x_1}{x_2 - x_1} \right)^{1 - 2 s}  \psi_s (x_1, \alpha x_2) - \alpha^{-1} \psi_s (x_1, x_2)
\right]
\, , \\
\widehat{H}^-_{12} \psi_s (\bit{x})
&
=
\int_0^1 \frac{d \alpha}{1 - \alpha} 
\
\left[
\left( \frac{x_2 - \alpha x_1}{x_2 - x_1} \right)^{1 - 2 s}  \psi_s (\alpha x_1, x_2) 
-
\psi_s (x_1, x_2) 
\right]
\, ,
\end{align}
taking on the form more resembling the boundary interactions due to the form of integration limits.

The same results can be obtained by using an intertwiner \cite{Belitsky:2014rba}
\begin{align}
W_N = (x_1 x_{21} \dots x_{N,N-1})^{2 s - 1}
\end{align}
that exchanges the index of the representation from $s$ to $1-s$. Namely,
\begin{align}
\widehat{H} = W H W^{-1}
\, .
\end{align}
We get
\begin{align}
\widehat{H}_{01} \psi_s (\bit{x})
&
=
\int_0^1 \frac{d \alpha}{1 - \alpha}
\left[
\left( \frac{x_2 - \alpha x_1}{x_2 - x_1} \right)^{1 - 2 s}  \psi_s (\alpha x_1, x_2) 
-
\psi_s (x_1, x_2) 
\right]
\, , \\
\widehat{H}_{2, \infty} \psi_s (\bit{x})
&
=
\int_1^\infty \frac{d \alpha}{\alpha - 1}
\left[
\left( \frac{\alpha x_2 - x_1}{x_2 - x_1} \right)^{1 - 2 s}  \psi_s (x_1, \alpha x_2) 
-
\alpha^{-1} \psi_s (x_1, x_2) 
\right]
\, , \\
[ \widehat{H}^+_{12} + \widehat{H}^-_{12} ] \psi_s(\bit{x})
&
=
\int_{x_1/x_2}^1 \frac{d \alpha}{1 - \alpha}
\left[
\psi_s (x_1, \alpha x_2) 
-
\psi_s (x_1, x_2) 
\right]
\\
&
+
\int_1^{x_2/x_1} \frac{d \alpha}{\alpha - 1}
\left[
\alpha^{1 - 2 s} \psi_s (\alpha x_1,  x_2) 
-
\alpha^{-1} \psi_s (x_1, x_2) 
\right]
\, . \nonumber
\end{align}
As we can see, the individual Hamiltonians do not jump places from bulk to boundary and back since the transformation mechanism is different compared to the 
integration by parts technique. The final cumulative answer is however the same as it has to be.

 %%%%%%  Bibliography %%%%%%%%%%%%%%%%%%%%%%%%%%%%%%%%%%%%%%%%%%%%

%%%%%%%%%%%%%%%%%%%%%%%%%%%%%%%%%%%%%%%%%%%%%%%%%%%%%%%%%%%%%%%%
\end{document}